\documentclass[10pt,english]{scrartcl}
\usepackage{amsmath}
\usepackage[T1]{fontenc}
\usepackage[ansinew]{inputenc}

\usepackage{babel}
\usepackage{graphicx}
\usepackage{amsmath}
\usepackage{amsfonts}
\usepackage{amssymb}
\usepackage{dsfont}
\usepackage[dvips]{color}
\newcommand{\vsO}{\vspace{.1cm}\hfill\\}

\newcommand{\vs}[1]{\vspace{#1mm}}
\title{Novel Analysis of Spinor Interactions\\and non-Riemannian Geometry} 

\author{{\normalsize O.M. Lecian$^{\;1}$, G. Montani$^{\;1,\,2}$, N. Carlevaro$^{\;1}$,}\vsO
\emph{\footnotesize $^1$ Physics Department, ``Sapienza'' University of Rome,}\vs{-2.5}\\
\emph{\footnotesize c/o Dipartimento di Fisica (VEF), P.le A. Moro 5 (00185) Roma, Italy}\vs{-2.5}\\
\emph{\footnotesize $^2$ ENEA -- UTFUS-MAG, C.R. Frascati (Rome, Italy)}\vs{-2.5}
}
\date{}

\begin{document}
\maketitle

{\bf   Abstract} A novel analysis of the gauge theory of the local Lorentz group is implemented both in flat and in curved space-time, and the resulting dynamics is analyzed in view of the geometrical interpretation of the gauge potential. The Yang-Mills picture of local Lorentz transformations is first approached in a second-order formalism. For the Lagrangian approach to reproduce the second Cartan structure equation as soon as the Lorentz gauge connections are identified with the contortion tensor, an interaction term between the Lorentz gauge fields and the spin connections has to be postulated. The full picture involving gravity, torsion and spinors is described by a coupled set of field equations, which allows one to interpret both gravitational spin connections and matter spin density as the source term for the Yang-Mills equations. The contortion tensor acquires a propagating character, because of its non-Abelian feature, and the pure contact interaction is restored in the limit of vanishing Lorentz connections. \qquad[PACS: 02.40.Hw, 11.15.-q, 04.62.+v]

\section{Introduction}\label{intro}

Gauge theories describe all physical interactions, but the gravitational one. Many attempts to construct a gauge model of gravitation exist, in particular the papers by Utiyama \cite{uti56} and by Kibble \cite{kib61} were the starting points for various gauge approaches to gravitation. As a result, Poincar\'{e} gauge theory (PGT) \cite{blago,hehvdhker76,heh73,heh74,hehvdhker74,hehvdhker74b,re4,deis,rel2,rel3} is a
generalization of the Einstein scheme of gravity, in which not only the energy-momentum tensor, but also the spin of matter plays a dynamical role when coupled to Lorentz connections, in a non-Riemannian space-time. Anyway, up to now, neither PGT nor other gauge approaches to the gravitational interaction have led to a consistent quantum scheme of the gravitational field. To include spinor fields consistently, it is necessary to extend the framework of General Relativity (GR), as already realized by Hehl \emph{et al.} \cite{hehvdhker76}: this necessity is strictly connected with the non existence in GR of an independent concept of spin angular momentum for physical fields, as the Lorentz group has not an independent status of gauge group in GR.

The Eintein-Cartan Theory (ECT) theory accounts for both mass and spin of matter as sources of the
gravitational field, and represent a description of gravity which is more suitable than GR from a microscopical point of view. In fact, fundamental interactions other than gravity are usually described within a theoretical framework, where matter is described by matter fields, and, spin, symmetries and conservation laws are properly encoded. In GR, contrastingly, matter can be described by point particles, fluids and light rays. This fundamental difference notwithstanding, spin effects are are negligible for macroscopic matter, so that the observational predictions of EC theory are regarded as the same as GR, from a phenomenological point of view \cite{re4}. Furthermore, ECT is a special case of PGT. However, PGT is much more general than ECT and encompasses also propagating Lorentz connections.

The gauge freedom of the gravitational interaction is the invariance under diffeomorphisms \cite{wal84}, strictly connected with the general-covariance principle. Moreover, the equivalence of any local reference system introduces another gauge freedom, mathematically expressed by the invariance of the model under local Lorentz transformations. But the former gauge freedom reabsorbs the latter, depriving the local Lorentz gauge group of independent gauge connections. Thus the dynamics of the gravitational field reduces to that of tetrads, while spin connections (introduced as gauge fields of local Lorentz transformations) have no longer an independent role in this framework, being only a particular combination of the tetrad fields and their derivatives.

The reasons of this have to be searched in the physical content of the two gauge transformations. From a mathematical point of view, using the tetrad formalism allows one to distinguish between diffeomorphisms and local Lorentz transformations, the first acting as a pull-back on the tetrad fields, while the second as a Lorentz rotation of the local basis; but, as we shall see, in the case of isometric coordinate transformations, the former can be restated in terms of the latter, and spin connections no longer have the proper change to ensure local Lorentz invariance. In fact, under isometric diffeomorphisms, spin connections transform as tensors, and cannot be gauge potentials for such diffeomorphism-induced local rotations.

Indeed, in the coordinate formalism, an infinitesimal diffeomorphism reads
\begin{equation}\label{inf diff} 
x^{\mu}\rightarrow x^{\prime \mu}=x^{\mu}+\alpha^{\mu}\left(x\right),
\end{equation}
where $\alpha^{\mu}\left(x\right)$ are four $C^{\infty}$ functions, while an
infinitesimal Lorentz (isometric) rotation has the form
\begin{equation}\label{inf diff-2}
x^{\mu}\rightarrow x^{\prime \mu}=x^{\mu}+\alpha^{\mu}_{\phantom1\nu}x^{\nu},
\end{equation}
where $\alpha^{\mu}_{\phantom1\nu}$ are the six infinitesimal rotational
parameters. For local Lorentz transformations, (\ref{inf diff-2}) can be reabsorbed in (\ref{inf diff}), i.e., 
\begin{equation}\label{inf loc lor}
x^{\mu}\rightarrow x^{\prime \mu}=x^{\mu}+\alpha^{\mu}_{\phantom1\nu}\left(
x\right)x^{\nu}=x^{\mu}+\tilde\alpha^{\mu}\left(x\right).
\end{equation}

This point of view is easily generalized to the case of scalar or macroscopic matter, which can be consistently approximated to spin-less matter, while gives rise to striking differences for spin-$\frac{1}{2}$ matter fields: fermions are described by a spinor representation of the Lorentz group, while the diffeomorphism group has no finite spinor representation. For the study of classical spinning particles interacting with the gravitational field, see also \cite{lus1}. The transformation law of fermions under local Lorentz rotations can no way be reabsorbed in the transformation law of tensor fields under the group of diffeomorphisms. Accordingly, the framework has to be further extended to consistently include fermions, giving back to the Lorentz group its status of independent gauge group \cite{hehvdhker76}, with its own connection fields, not directly related with the tetrad fields, as they are in General Relativity.

Nevertheless, it is worth stressing that infinite representations of such a group do exist, as shown in \cite{neem,neem1} and the references therein. Finite spinors can introduced either by making use of the non-linear representations of the double covering of the general-coordinate-transformation group, which are linear when restricted to the Poincar\'{e} subgroup, or by introducing a bundle of cotangent frames and defining in this space the action of a physically-distnct Lorentz group. In the second case, after generalizing the Lorentz group, infinite-dimensional linear spinor representations or finite-dimensional non-linear spinor representations can be found. In \cite{neem1}, infinite-component spinor and tensor fields (so-called manifields) are introduced: these manifields are are then lifted to the proper corresponding representation via the introduction of infinite-component frame-fields.

Since, in our approach, the introduction of a Lorentz gauge field is based on the fact that spinors behave as a representation of the Lorentz group \cite{g2}, translations (i.e., non-isometric infinitesimal diffeomorphisms) are not included in this gauge picture, because, from this point of view, translations are not distinguishable from generic diffeomorphisms.

In this respect, it is worth recalling that the teleparallel theory of gravity can be treated physically as a gauge theory of translations. In fact, teleparallel gravity can be understood within the framework of metric-affine gravitational theories \cite{mag2}, and it is picked up from such other models by reducing the affine symmetry group to the translation subgroup, i.e. by imposing vanishing curvature and non-metricity. Within the framework of a metric-affine approach to teleparallel gravity, the introduction of spinless matter, characterized only by the energy-momentum, can be illustrated to be completely consistent with teleparallelism, while the case of spinning matter sources exhibits a consistency problem, and teleparallelism appears to be not applicable in the second case \cite{obu}. On the other hand, formulations have been proposed, in which gravity can be understood as deriving from translation symmetry (see, for example \cite{obu,gag,hehlmac} and all the references therein).

PGT is the theoretical framework that enables one to take into account both translations and Lorentz transformations, and spinor dynamics in curved space-time is recovered by the introduction of compensating fields that restore local invariance \cite{blago}. The present proposal \cite{g2,naka,nakb} differs from PGT also because we obtain differential dynamical equations for the gauge connections, rather than algebraic ones. Nevertheless, a contact interaction is recovered in the limit of vanishing Lorentz connections, where the relation between the spin connections and the spin density becomes algebraic. For another propagative approach to torsion, see, for example, \cite{g1a,g1b,g1c}.
 
The paper is organized as follows. The features of GR that make it resemble to a gauge theory are reviewed in the first Section, where the tools which will support our analysis, i.e., the tetrad formalism and Cartan structure equations, are revised, and the properties of local Lorentz transformations are outlined.
The second Section is aimed at describing PGT both from  gauge and from  geometrical points of view. The geometrical structure of PGT and the geometrical meaning of gauge fields and conserved quantities will follow from the comparison. This section is mostly devoted at stressing those features that allow us to examine differences and consider similarities with the Lorentz-gauge proposal.
In Section 3, we will briefly revise the main features of Teleparallel geometry, understood as a limit of PGT, and particular attention will be devoted to the case of parallelizable manifolds. In fact, in this case, the interpretation of the scheme as a gauge theory of translations can be conceived as complementary of a gauge theory of the Lorentz group. 
In Section 4, we will show the possibility to restate isometric diffeomorphisms in terms of local Lorentz rotations, both in their finite and infinitesimal character.
A gauge theory for the local Lorentz group on flat space-time is implemented in Section five. After defining the space where local Lorentz transformations take place, we will implement Lorentz gauge transformations for spinor fields, from which the non-Abelian character of the gauge field will be inferred, and a suitable bosonic Lagrangian density will be established, accordingly. Finally, Dirac and Yang-Mills equations will be derived, where the interaction between spinors and the gauge fields shows up \cite{g2,g1a,g1b,g1c,naka,nakb}.
The results of the previous section will be generalized to curved space-time in the sixth Section. Local Lorentz transformations in curved space-time will be analyzed, and two different approaches (second- and first-order formalism, respectively) to implement a Lorentz gauge theory in such a scenario will be followed. Dirac, Yang-Mills, and Einstein equations will be derived in both cases. In particular, the relation between suitable bein projections of the contortion field and Lorentz gauge fields will be hinted in the second-order formalism, and the geometrical hypotheses for such an identification will be investigated in the first-order formalism. An interaction term between the Lorentz gauge field and the spin connections has to be postulated in order to restore the proper mathematical identification from the second Cartan structure equation. First- and second-order approaches will be eventually compared in the linearized regime. In Section 7, brief concluding remarks follow.\vspace{5mm}

{\footnotesize Throughout this paper, we adopt the following notation:
\begin{itemize}
  \item lower-case Greek letters from the beginning of the alphabet (i.e. $\alpha$, $\beta$) denote 4-dimensional inertial indices, $\alpha=0,1,2,3$;
	\item lower-case Greek letters from the middle of the alphabet (i.e. $\mu$, $\nu$) denote general 4-dimensional world indices, $\mu=0,1,2,3$;
	\item lower-case Latin letters from the beginning of the alphabet (i.e. $a$, $b$) denote 4-dimensional bein indices, $a=0,1,2,3$;
	\item lower-case Latin letters from the middle of the alphabet (i.e. $m$, $n$) denote the rank of tensors or the dimension of spaces;
	\item lower-case Latin letters from the second half of the alphabet (i.e. $r$, $s$) denote spinor indices, $r=1,2$;
	\item repeated indices are summed over for tensor objects and the summation symbol is usually omitted (i.e. $v^\mu v_\mu\equiv\sum_{\mu=0}^{\mu=3}v^\mu v_\mu$)
	\item lower-case Latin letters from the first part of the alphabet (i.e. $i$, $j$) denote indices whithin summation, to which peculiar attention attention has to be devoted;
  \item $\wedge$ denotes exterior product;
  \item $\star$ denotes the Hodge operator for forms ( i.e. $\star\eta=\frac{1}{p!}\eta_{m_1...m_p}\epsilon^{m_1...m_p}_{\ \ \ \ \ \ \ \  m_{p+1}...m_{n}}dx^{m_{p+1}}\wedge...\wedge dx^{m_n}$, where $\eta$ is a p-form on a n-dimensional manifold);
  \item $\bar{}$ (overbar) denotes adjoint spinors (i.e. $\bar{\psi}=\psi^\dag\gamma^0$, where $\psi$ is a spinor, $\psi^\dag$ the conjugate spinor and $\gamma^0$ the zero-th Dirac matrix).
\end{itemize}}


\section{General Relativity as a gauge theory}\label{par1}

This section is aimed at describing GR as a gauge model, i.e., at pointing out the features that render the comparison possible, as well as those that make this theory isolated form the gauge picture, such as the metric field, which has no analogues in other known physical interactions.
  
\paragraph{Metric tensor and tetrad field}\null
 Let $M^{4}$ be a 4-dimensional pseudo-Riemannian manifold , and $e$ a one-to-one map on it, $e:M^{4}\rightarrow TM^{4}_{x}$, which
sends tensor fields on $M^{4}$ in tensor fields in the Minkowskian tangent space
$TM^{4}_{x}$: the fields $e^{\phantom1a}_{\mu}$ are called tetrads or vierbein, and are an
orthonormal basis for the local Minkowskian space-time; Greek indices ($\mu=0,1,2,3$) change as tensor ones under general coordinate
transformations, while Latin indices ($i=0,1,2,3$) refer to local Lorentz transformations. The following relations between the tetrads and the metric field $g_{\mu\nu}$ of the manifold $M^{4}$ hold 
\begin{equation}\label{relation}
g_{\mu\nu}=\eta_{ab}e_{\mu}^{\phantom1a}e_{\nu}^{\phantom1b},\quad
e_{\mu}^{\phantom1a}e^{\mu}_{\phantom1b}=\delta^{a}_{b},\quad
e_{\mu}^{\phantom1a}e^{\nu}_{\phantom1a}=\delta^{\nu}_{\mu},
\end{equation}
where $\eta_{ab}$ is the metric tensor in the local Minkowski frame. Eq. (\ref{relation}) states that, given the tetrad field $e^{\phantom1a}_{\mu}$, the metric tensor $g_{\mu\nu}$ is uniquely determined, and all metric properties of the space-time are expressed by the tetrad field, accordingly. It is worth noting that the converse is not true:
there are infinitely many choices of the local basis that reproduce the same
metric tensor, because of the local Lorentz gauge invariance. This is also the
reason why there are more components in $e_{\mu}^{\phantom1a}$ than in the
metric $g_{\mu\nu}$, the difference being six, i.e., the number of
independent rotations in four dimensions.

Rewriting usual tensor equations in a frame formalism allows us to project tensor fields from the 4-dimensional manifold to the Minkowskian space-time, thus emphasizing the local Lorentz invariance of the scheme. Moreover, to assure that the projected derivative of a tensor field be invariant under local Lorentz transformations, the connection 1-forms $\omega^{a}_{\phantom1b}$ must be introduced, which take values in the adjoint representation of the Lorentz group. They define the covariant exterior derivative operator $d^{(\omega)}$, whose action on Lorentz-valued differential forms is
\begin{align}
\nonumber d^{(\omega)}v^{a_1,\dots,a_n}_{c_{1},\dots,c_m}=d
v^{a_1,\dots,a_n}_{c_{1},\dots,c_m}+ &
\Sigma^{n}_{i=1}\omega^{a_i}_{\phantom1b_i}\wedge
v^{a_1,\dots,b_i,\dots,a_n}_{c_{1},\dots,c_m}+
\\
- & \Sigma^{m}_{j=1}\omega^{d_j}_{\phantom1c_j}\wedge
v^{a_1,\dots,a_n}_{c_1,\dots,d_j,\dots,c_m}.
\end{align}
For later purposes, it's worth remarking that the introduction of the tetrad formalism enables one to include spinor fields in the dynamics, as spinor fields are a particular representation of the Lorentz group.

\paragraph{Structure equations}\null
Connection 1-forms lead to the usual definition of the
curvature 2-form $R^{a}_{\phantom1b}$:
\begin{equation}\label{Riemann}
d^{(\omega)}\circ d^{(\omega)}v^a=R^{a}_{\phantom1b}\wedge v^b,
\end{equation}
or
\begin{equation}
R^{a}_{\phantom1b}=d\omega^{a}_{\phantom1b}+\omega^{a}_{\phantom1c}\wedge\omega^{c}_{\phantom1b},
\end{equation}
which is the first Cartan structure equation, and can be rewritten as the Bianchi
identity
\begin{equation}\label{bianchi1}
d^{(\omega)}R^{a}_{\phantom1b}=0.
\end{equation}
In this formalism, the action for GR consists of the lowest-
order non-trivial scalar combination of the Riemann curvature 2-form
(\ref{Riemann}) and the tetrad fields, that is the Hilbert-Einstein
action
\begin{equation}\label{action for o}
S(e,\omega)=\frac{1}{4}\int\epsilon_{a b c d}\,e^{a}\wedge e^{b}\wedge R^{c d}.
\end{equation}
Variation with respect to connections leads to the
second Cartan structure equation in the torsion-less case,
\begin{equation} \label{Cartan eq}
d e^{a}+\omega^{a}_{\phantom1b}\wedge e^{b}=0,
\end{equation}
which links the tetrad fields to the spin connections and leads to the identity
\begin{equation} \label{cyclic identity}
e^{b}\wedge R^{a}_{\phantom1b}=0.
\end{equation}
Variation with respect to tetrads leads to the
following equations:
\begin{equation}
\epsilon_{a b c d}\,e^{b}\wedge R^{c d}=0,
\end{equation}
which give the dynamical Einstein field equations, once the solution of the second Cartan structure equation (\ref{Cartan eq}) is considered.

We remark that, under local Lorentz transformations, spin-connections transform like a Lorentz gauge vector, and the Riemann 2-form is preserved by such a change. Therefore, in flat space-time, we deal with non-zero spin connections, but a vanishing curvature 2-form.
In both flat and curved space-time, spin connections exhibit the right behavior to play the role of Lorentz gauge fields, and GR has the features of a gauge theory, despite some relevant shortcomings outlined below.

\paragraph{Lie Derivatives and Killing Vectors}
A coordinate transformation $x\rightarrow x'$ that does not modify the form of the metric defines the isometry group of a given space. For the infinitesimal transformation $x^\mu\rightarrow x'^\mu=x^\mu+\xi^\mu$, the invariance of the metric is expressed by the Killing equation $\xi_{\mu,\nu}+\xi_{\nu,\mu}=0$. Expanding $\xi^\mu$ in power series of $x$, we find that global Poincar\'{e} transformations have the form $\xi^\mu=\epsilon^\mu+\omega^\mu_\nu x^\mu$, and are defined in terms of ten constant parameters $\omega^{\mu\nu}=-\omega^{\nu\mu}$ and $\epsilon^\mu$, i.e. these transformations are translations and Lorentz rotations.

It is possible to associate a vector field to a one-parameter family of diffeomorphisms \cite{wal84}. In fact, given a manifold $M$, a one-parameter group of diffeomorphisms $\phi_t$ is a $C^\infty$ map from $\mathbb{R}\times M\rightarrow M$ such that for any fixed $t\in\mathbb{R}$, $\phi_t:M\rightarrow M$ is a diffeomorphism, $\phi_t\circ\phi_s=\phi_{t+s}$, and $\phi_{t=0}$ is the identity. Then for any point $p\in M, \phi_t(P):\mathbb{R}\rightarrow M$, there is a curve that passes through $p$ at $t=0$. This way, $v_p$ is the tangent to this curve at $t=0$. This way, the vector field $v$ is associated to the one-parameter group of transformations, and can be regarded as the generator of these transformations.

Given a manifold $N$ and a function $f$ on it, $f:N\rightarrow\mathbb{R}$, such that $f\circ\phi:M\rightarrow\mathbb{R}$, $\phi^*$ is defined as $(\phi^* v)(f)=v(f\circ\phi)$.
The Lie derivative $\mathcal{L}_v$ of a smooth tensor field $T^{a_1...a_m}_{b_1...b_n}$ at the point $p$ with respect to $v$ is defined as
\begin{equation}
\mathcal{L}_vT^{a_1...a_m}_{b_1...b_n}=\lim_{t\rightarrow0}\frac{\phi^*_{-t}T^{a_1...a_m}_{b_1...b_n}-T^{a_1...a_m}_{b_1...b_n}}{t},
\end{equation}  
and is a linear map from tensors of rank $(m,n)$ to tensors of rank $(m,n)$. Furthermore, $\mathcal{L}_vT^{a_1...a_m}_{b_1...b_n}=0$ everywhere iff $\phi_t$ is a symmetry transformation for $T^{a_1...a_m}_{b_1...b_n}$ for all $t$.

When specified for the features of the gravitational field, the discussion above leads to the definition of isometries for the metric tensor. In fact, if $\phi_t:M\rightarrow M$ is a one-parameter group of isometries, $\phi^*_t g_{\mu\nu}=g_{\mu\nu}$, the vector field $\xi$ that generates $\phi_t$ is a Killing vector field, and $\xi$ satisfies the Killing equation $\nabla_\mu\xi_\nu+\nabla_\nu\xi_\mu=0$.

Despite the use of the Lie derivative is the most appropriate and synthetic approach to the characterization of the properties of the local translation group, in what follows we will retain the  usual space-time covariant formulation in order to keep  the mathematical formulation of the gauge theory of the Lorentz group as close as possible to its physical implementation at the level of field dynamics, in agreement with common approaches \cite{wal84}.

\paragraph{Physical interpretation }\null
So far, it is possible, on the one hand, to appreciate the similarity with gauge theories, the role of connection fields being played by spin connections, $\omega_{\mu\phantom1b}^{\phantom1a}$, and, on the other hand, to remark that the presence of the tetrad field, introduced by the Principle of General Covariance, is an ambiguous element for a gauge paradigm. This scenario would be appropriate if the theory were based on two independent degrees of freedom. Since spin connections can be uniquely determined as functions of tetrad fields, (\ref{Cartan eq}), then this correlation opens a puzzle in the interpretation of these connections as the only fundamental fields of the gauge scheme.

Nonetheless, it is just the introduction of fermions that requires to treat local Lorentz transformations as the real independent gauge of GR. In fact, when spinor fields are taken into account, their transformations under the local Lorentz symmetry imply that the Dirac equation is endowed with non-zero spin connections even on Minkowskian space-time, as developed in section $5$.
Because of the behavior of spinor fields, it becomes crucial to investigate whether diffeomorphisms can be reinterpreted to some extent as local Lorentz transformations.

\section{Poincar\'{e} gauge theory}
In this section, we analyze a well-established proposal to connect the presence of torsion with the local nature of the Poincare\'{e} symmetry. This approach represents a comparison scheme for the present developments of the model outlined here. PGT will be described from both a gauge and a geometrical point of view, and particular attention will be payed to the physical meaning of field equations, which predict a contact interaction, i.e., a non-propagating gauge field.

\paragraph{Gauge approach}\null
Let us consider an infinitesimal global Poincar\'{e} transformation in Minkowski space,
\begin{equation}\label{poincare global}
x^{\mu}\rightarrow x^{\prime \mu}=x^{\mu}+\tilde{\epsilon}^{\mu}_{\phantom1\nu}x^{\nu}+\tilde{\epsilon}^{\mu},
\end{equation}
and the consequent transformation law for spinor fields
\begin{equation}\label{new poincare global}
\psi\left(x\right)\rightarrow\psi^{\prime}\left(x\right)=\left(1+\frac{1}{2}\tilde{\epsilon}^{\mu\nu}M_{\mu\nu}+\tilde{\epsilon}^{\mu}P_{\mu}\right)\psi\left(x\right),
\end{equation}
where the generators $M_{\mu\nu}=L_{\mu\nu}+\Sigma_{\mu\nu}$ and $P_{\mu}$ obey Lie-algebra commutation relations. If the matter Lagrangian density is assumed to depend on the spinor field and on its derivatives only, $L=L(\psi, \partial_{\mu}\psi)$, and if the equations of motion are assumed to hold, the conservation law $\partial_{\mu}J^{\mu}=0$ is found, where
\begin{equation}\label{j}
J^{\mu}=\frac{1}{2}\tilde{\epsilon}^{\nu\lambda}M^{\mu}_{\ \nu\lambda}-\tilde{\epsilon}^{\nu}T^{\mu}_{\ \nu},
\end{equation}
where the canonical energy-momentum and angular-momentum tensors are defined, respectively, as
\begin{equation}
T^{\mu}_{\ \nu}=\frac{\partial L}{\partial \psi,_{\mu}}\partial_{\nu}\psi-\delta^{\mu}_{\ \nu}L,
\end{equation}
\begin{align}
M^{\mu}_{\ \nu\lambda}=& \left(x_{\nu}T^{\mu}_{\ \lambda}-x^{\lambda}T^{\mu}_{\ \nu}\right)-S^{\mu}_{\ \nu\lambda}\equiv \nonumber
\\
& \equiv \left(x_{\nu}T^{\mu}_{\ \lambda}-x^{\lambda}T^{\mu}_{\ \nu}\right)+\frac{\partial L}{\partial \psi,_{\mu}}\Sigma_{\nu\lambda}\psi.
\end{align}
Because the parameters in (\ref{j}) are constant, according to the Noether theorem, conservation laws for the energy-momentum current and for the angular-momentum current, together with the related charges, are established\footnote{if the integration on the boundaries of the 3-space brings vanishing contributions.}:
\begin{equation}
\partial_{\mu}T^{\mu}_{\ \nu}=0\Rightarrow P^{\nu}=\int d^{3}x T^{0\nu}
\end{equation}
\begin{equation}\label{m}
\partial_{\mu}M^{\mu}_{\ \nu\lambda}=0\Rightarrow M_{\nu\lambda}=\int d^{3}xM^{0}_{\nu\lambda}.
\end{equation}
When transformations are locally implemented, eq.s (\ref{j})-(\ref{m}) do not hold any more, and compensating gauge fields have to be introduced in order to restore local invariance. As a first step, a covariant derivative $D_{a}\psi$ is defined as 
\begin{equation}\label{derivatives}
D_{a}\psi=e_{a}^{\ \mu}D_{\mu}\psi=e_{a}^{\ \mu}\left(\partial_{\mu}+A_{\mu}\right)\psi=e_{a}^{\ \mu}\left(\partial_{\mu}+\frac{1}{2}A_{\mu}^{ab}\Sigma_{ab}\right)\psi,
\end{equation}
where the compensating fields $e^{k}_{\ \mu}$ and $A_{\mu}^{ij}$, and the generator $\Sigma_{ij}$ have been taken into account. This way, the Lagrangian density depends on the covariant derivative of the fields, instead of the ordinary one, $L=L(\psi,D_{a}\psi)$; covariant derivatives (\ref{derivatives}) do not commute, but satisfy the commutation relation
\begin{equation}\nonumber
[D_{\mu},D_{\nu}]\psi=\frac{1}{2}F^{ab}_{\ \ \mu\nu}\Sigma_{ab}\psi,
\end{equation}
\begin{equation}
[D_{d},D_{f}]\psi=\frac{1}{2}F^{ab}_{\ \ df}\Sigma_{ab}\psi-F^{c}_{\ df}D_{c}\psi,
\end{equation}
where $F^{ab}_{\ \ \mu\nu}$ and $F^{c}_{\ df}$ are the Lorentz field strength and the translation field strength, respectively.

Covariant energy-momentum and spin currents, $T'^{\mu}_{\ \nu}$ and $S'^{\mu}_{\  ab}$, can be found, in analogy with the global case, after the substitution $\partial_{\mu}\rightarrow D_{\mu}$, and are found to be equivalent, if the equation of motion for matter fields are assumed to hold, to the dynamical currents $\tau^{\mu}_{\ \nu}$ and $\sigma^{\mu}_{\ \ ab}$,
\begin{equation}
T'^{\mu}_{\ \nu}=\frac{\partial L}{\partial D_{\mu}\psi}D_{\nu}\psi-\delta^{\mu}_{\ \nu}L\equiv\tau^{\mu}_{\ \nu}=e_{a}^{\ \mu}\frac{\partial L}{\partial e_{a}^{\ \nu}},
\end{equation}
\begin{equation}\label{spin}
S'^{\mu}_{\ ab}=-\frac{\partial L}{\partial D_{\mu}\psi}\Sigma_{ab}\psi\equiv\sigma^{\mu}_{\ ab}=-\frac{\partial L}{\partial A^{ab}_{\ \ \mu}}.
\end{equation}
As outlined in \cite{hehvdhker76}, it is possible to infer the inadequacy of special relativity to describe the behavior of matter fields under local Poincar\'{e} transformations. Global Poincar\'{e} transformations preserve distances between events and the metric properties of neighboring matter fields: comparing field amplitudes before performing the transformation, and then transforming the result, or comparing the transformed amplitudes of the fields is equivalent. This property is known as rigidity condition, as matter fields behave as rigid bodies under this kind of transformations. On the contrary, it can be shown that the action of local Poincar\'{e} transformations can be interpreted as an irregular deformation of matter fields, thus predicting different phenomenological evidences for the field and for the transformed field.
The compensating gauge fields $e_{\mu}^{c}$ and $A^{ab}_{\ \ \mu}$, introduced to restore local invariance, describe geometrical properties of the space-time, as it can be easily argued from a geometrical point of view.

\paragraph{Geometrical approach}\null
The geometrical approach to PGT can be carried out by considering the most general metric-compatible linear connections,
\begin{equation}\label{cane}
\Gamma^{\mu}_{\nu\rho}=\left\{
\begin{array}{c}
\mu\\
\nu\rho
\end{array}
\right\}-K_{\ \nu\rho}^{\mu},
\end{equation}
where $\left\{
\begin{array}{c}
\mu\\
\nu\rho
\end{array}
\right\}
\equiv
\left\{
\begin{array}{c}
\mu\\
\rho\nu
\end{array}
\right\}$ are the Christoffel symbols, and $K^{\mu}_{\ \nu\rho}\equiv K^{\mu}_{\ \rho\nu}$ the contortion tensor, with 24 independent components, which can be written as a function of the torsion field $T^{\mu}_{\ \nu\rho}$,
\begin{equation}
K^{\mu}_{\ \nu\rho}=-\frac{1}{2}\left(T^{\mu}_{\ \nu\rho}-T_{\rho\ \nu}^{\ \mu}+T^{\ \ \mu}_{\nu\rho}\right).
\end{equation}
Geometric covariant derivatives are defined as
\begin{equation}
D_{\mu}\psi=\left(\partial_{\mu}+\omega_{\mu}\right)\psi=\left(\partial_{\mu}+\frac{1}{2}\omega^{ab}_{\ \ \mu}\Sigma_{ab}\right)\psi,
\end{equation}
where the spin connections $\omega^{ab}_{\ \ \mu}$ consist of the bein projections of the Ricci rotation coefficients and the contortion field, respectively: $\omega_{ab\mu}=R_{ab\mu}+K_{ab\mu}$.

The gauge potentials $e^{\phantom1\mu}_a$ are generally interpreted as the relation between the orthonormal frame (Greek indices) and the coordinate one (Latin indices), while the introduction of the gauge potentials $\omega_\mu^{ab}$ is connected with the corresponding rotations of the orthonormal basis at neighboring points: this induces a change in the derivative operator, i.e.
\begin{equation}
\partial_\mu\rightarrow D_\mu\equiv\partial_\mu+\frac{1}{2}\omega_\mu^{ab}\Sigma_{ab}.
\end{equation}
Torsion contributes to the gravitational dynamics, according to its gravitational action: it has been illustrated \cite{hashi} that the most general form for a Lagrangian $L_{T}$ (which allows for equation of motion that are at most of second order in the field derivatives) is
\begin{equation}\label{lagtor}
L_{T}=AT_{abc}T^{abc}+BT_{abc}T^{bac}+CT_{a}T^{a}\equiv b_{abc}T^{abc},
\end{equation}
where $T_{a}=T^{b}_{\ ba}$ and $b_{abc}=a\bigl(AT_{abc}+BT_{[bac]}+C\eta_{a[b}T_{c]}\bigr)$. The values of the parameters $A,B,C$ are to be determined according to the Physics that has to be described, and some relevant examples are discussed in \cite{rel1,rel2,rel3,re4}.

\paragraph{Discussion}\null
The comparison of gauge and geometrical approaches leads to the identification of the Lorentz gauge fields $A^{ab}_{\ \mu}$, which accounts for local Lorentz transformations, with the spin connections $\omega^{ab}_{\ \ \mu}$, and the fields $e_{a}^{\ \mu}$, which describe translations, with the components of the tetrad field. This way,  the identifications of the Lorentz field strength with curvature, and that of the translation field strength with torsion, are straightforward.

After the introduction of covariant derivatives, the coordinate representation of the generators of translations, $P_{\mu}\rightarrow\partial_{\mu}$, changes as $P_{\mu}\rightarrow D_{\mu}$ with the implementation of local transformations; the variation of matter fields, in the two cases, differs by a local Lorentz rotation, thus mixing up the concept of translations and Lorentz rotations \cite{tietre05-2}.

The analysis of \cite{weisa,weisb}, based on the conception of Poincar\'{e} symmetry as a purely inner symmetry on Minkowskian background, sheds light on the relation between gravity and the structure of spacetime. According to this scheme, the global action of the Poincar\'{e} group is analogous to the description of the action of inner-symmetry groups as groups of generalized rotations in the field space. The interpretation of the resulting theory as a gauge theory of gravitation is achieved by requiring that the corresponding local invariance show up as invariance under general coordinate transformations and local $SO(3,1)$ frame rotations. 

Furthermore, the usual Einstein-Hilbert action for gravity in four dimensions is not invariant under the translational part of the Poincar\'{e} group, while is invariant under the Lorentz group.

Because of these unpleasant features, the straightforward interpretation of PGT as a gauge theory of gravity implies a non-trivial modification of the geometrodynamics.
Field equations read:
\begin{subequations}
\begin{align}
&\frac{1}{e}D_\mu\left(e e^\nu_{a}e^\mu_{b}\right)=S_{\phantom1 ab}^{\nu},
\\
&R_{ab}^{\phantom1\phantom1 \mu b}-\frac{1}{2}\,e^\mu_{\phantom1 a}R_{cd}^{cd}=T_{a}^\mu.
\end{align}
\end{subequations}
The first equation provides the expression of the connections of the rotation group as a function of the connections of the translation group and the matter fields, while the second equation is the Einstein dynamical equation: tensor fields involved in the equations above must satisfy identities (\ref{bianchi1}) and (\ref{cyclic identity}). By the geometrical identification of covariant gauge derivatives, (\ref{spin}) becomes an algebraic relation between spin and torsion: since the relation is not differential, torsion is not predicted to propagate, but its existence is bound to the presence of spin-$\frac{1}{2}$ matter fields, i.e., torsion is not a real dynamical field.

For later convenience, it will be useful to restate the description of PGT in a slightly different formalism, which allows for a better explanation of the role of spin, so that the analogies and differences with the Lorentz gauge theory will be more noticeable.
Eq. (\ref{new poincare global}) can be written as 
\begin{equation}\label{new poincare global 2}
\psi\left(x\right)\rightarrow\psi^{\prime}\left(x\right)=\left(1+\frac{1}{2}\epsilon^{\mu\nu}\Sigma_{\mu\nu}+\epsilon^{\mu}P_{\mu}\right)\psi\left(x\right), 
\end{equation}
where $\epsilon^{\mu}\equiv\widetilde{\epsilon}^{\mu}+\widetilde{\epsilon}_{\nu}^{\phantom1\mu}\delta^{\nu}_\rho x^\rho$, $\epsilon^{\alpha\beta}=\widetilde{\epsilon}^{\alpha\beta}$, and the generators of translations and spin rotations satisfy the relations:
\begin{align}
& \left[\Sigma_{ab},\Sigma_{cd}\right]=\eta_{c[a}\Sigma_{b]d}-\eta_{d[a}\Sigma_{b]c},\nonumber
\\
& \left[\Sigma_{ab},P_{c}\right]=-\eta_{c[a}\partial_{b]},\quad
 \left[P_{a},P_{b}\right]=0.
\end{align}
The advantage of (\ref{new poincare global 2}) consists in keeping pure rotations separated from translations. The orbital angular momentum is this way kept independent of the spin angular momentum: the former is strictly related with the energy-momentum, thus with the rotation-dependent part of $\epsilon^{\mu}$, while the latter is connected with the pure-rotation parameter $\epsilon^{\mu\nu}$. In fact, if the analogy is drawn between a generic diffeomorphism and a global Poincar\'{e} transformation, it is impossible to perform translations and rotations independently, but, when a localized symmetry is considered, this becomes possible, because the parameters defining the transformation are allowed to vary freely.

\section{Teleparallelism}
An interesting limit of PGT is Weitzenb\"ock or  
teleparallel geometry , defined by the requirement
\begin{equation}
R^{ab}{}_{\mu\nu}(A)=0.                                   \label{3.1}
\end{equation}
Teleparallel geometry (see, for example, \cite{re4} for a hand-on review and all the references therein) can be interpreted, to some extents, as complementary to Riemannian: curvature vanishes and torsion
remains to characterize the parallel transport. The physical interpretation of such a geoemtry relies on the fact that there is a one-parameter family of teleparallel Lagrangians which is empirically equivalent to GR
\cite{13,14,15}.

\paragraph{Lagrangian and Field equations} Within this framework, the gravitational field is described by tetrads $e^a{_\mu}$ and Lorentz
connections $A^{ab}{_\mu}$, where (\ref{3.1}) has to be taken into account. For our purposes, it is usefull to consider the set of Lagrangians, which are quadratic with respect to the torsion fiel, i.e.
\begin{equation}
L_{TP}=bL_T +\lambda_{ab}{}^{\mu\nu}R^{ab}{}_{\mu\nu}+L_M,                       \label{3.2}
\end{equation}
where $\lambda_{ab}{}^{\mu\nu}$ are Lagrange multipliers which imply requirement (\ref{3.1}) in the variational formalism,
and $L_T$ has been defined in (\ref{lagtor}). Variation of (\ref{3.2}) with
respect to the object $e^a{_\mu}, A^{ab}{_\mu}$ and $\lambda_{ab}{}^{\mu\nu}$ lead to the following field equations:
\begin{subequations}\label{3.5}
\begin{align}
4\nabla_\rho(b\beta_a{^{\mu\rho}})-4b\beta^{bc\mu}T_{bca}
     +e_a{^\mu}bL_T &=\tau^\mu_{a}\, ,                       \label{3.5a}\\
4\nabla_\rho\lambda_{ab}{}^{\mu\rho}
     -8b\beta_{[ab]}{^\mu} &=\sigma^\mu{}_{ab} \, ,                  \label{3.5b}\\
R^{ab}{}_{\mu\nu} &=0 \, .                                  \label{3.5c}
\end{align}
\end{subequations}

Equation (\ref{3.5c}) ensures (\ref{3.1}) from variational principles. Equation (\ref{3.5a}) characterizes $e^a{_\mu}$ from a dynamical point of view.
Eq. (\ref{3.5b}) determines the Lagrange multipliers $\lambda_{ab}{}^{\mu\nu}$, and the non-uniqueness of $\lambda_{ab}{}^{\mu\nu}$ is
related to an extra gauge freedom in the theory. In fact,
the gravitational Lagrangian (\ref{3.2})
is, by construction, invariant under the local Poincar\'{e}
transformations, and, up to a
four-divergence, under the set of transformations \cite{16} 
\begin{equation}
\delta\lambda_{ab}{}^{\mu\nu}=\nabla_\rho\varepsilon_{ab}{}^{\mu\nu\rho}\;,\qquad
\delta\lambda_{ab}{}^{jk}=\nabla_0\varepsilon_{ab}{}^{jk}+\nabla_i\varepsilon_{ab}{}^{jki}\;,\qquad
\delta\lambda_{ab}{}^{0j}=\nabla_i\varepsilon_{ab}{}^{ji}\;,\label{3.7a}
\end{equation}
where $\varepsilon_{ab}{}^{\mu\nu\rho}\equiv\varepsilon_{ab}{}^{[\mu\nu\rho]}$ and $\varepsilon_{ab}{}^{\mu\nu}\equiv\varepsilon_{ab}{}^{\mu\nu 0}$: this way, Eq.(\ref{3.1}) implies Eq.(\ref{3.5b}).
 
The $\lambda$ symmetry is defined by the
$\varepsilon_{ab}{}^{\mu\nu}$ parameters, so that
the six parameters $\varepsilon_{ab}{}^{\mu\nu i}$ may be removed by defining $18$ $\lambda_{ab}{}^{\mu\nu}$, while the other $18$ degrees of freedom are given by Eq. (\ref{3.5b}). The gauge structure of such a one-parameter teleparallel theory is believed to be still problematic \cite{17a,17b,17c} 

\paragraph{Orthonormal frames} For a parallelizable manifold, the path-independence of parallel transport is implied by (\ref{3.1}); an orthonormal frame is defined by vanishing connections:
\begin{align}\label{3.8}
A^{ab}{_\mu}=0\;.
\end{align}
This solution is not uniquely defined, but global Lorentz relate each class of orthonormal frames obtained. Because of this, the covariant derivatives become partial derivatives, with $T^a{}_{\mu\nu}=\partial_\mu e^a{_\nu}-\partial_\nu e^a{_\mu}$. Because of the action of Lorentz transformations of the tetrad fields on the connection, i.e.
$$
e'{^a}{_\mu}=\Lambda^a{_c}e^c{_\mu} \quad \Rightarrow \quad
A'^{ab}{_\mu}=\Lambda^a{_c}\Lambda^b{_d}A^{cd}{_\mu}+\Lambda^a{_c}\partial_\mu\Lambda^{bc}\, ,
$$
the solution of $R^{ab}{}_{\mu\nu}(A)=0$ is given by $A^{ab}{_\mu}=\Lambda^a{_c}\partial_\mu \Lambda^{bc}$: this way, the definition of (\ref{3.8})
deos not respect local Lorentz invariance, and is interpreted as a definition that fixes the gauge.

 \paragraph{Discussion}
In (\ref{3.2}), teleparalllism is given by the Lagrange multipliers $\lambda$, defined by (\ref{3.5b}), while (\ref{3.5a}) accounts for the non-trivial features of the dynamics, such that, for a parallelizable manifold, imposing (\ref{3.8}) in the action results in teleparallelism, interpreted as a gauge field of translations given uniquely by the tetrad field.

The consistency of teleparallel gravity when spinning matter is taken into account has also been discussed within the framework of the teleparallel limit of PGT \cite{lecl,lecl2}. In \cite{lecl}, an inconsistency, due to frame dependence, was illustrated to arise for every gauge theory of the Poincar\'{e} group that admits a teleparallel limit in the absence of spinning matter. Furthermore, in \cite{lecl2}, a restricted class of transformations was found, according to which the frame invariance of the gravitational Lagrangian does not lead to inconsistencies, even as far as Standard-Model particles are concerned, and experimental aspects were analyzed.

\section{Motivation for a gauge theory of the Lorentz group}
In what follows, we will discuss separately the two cases of flat and curved spacetime as far as the implementation of the geometrical gauge proposal for the Lorentz symmetry is concerned. Here, we wish to fix some key points, which are at the ground of the physical motivation for such a Lorentz gauge theory as a non-Riemannian effect. The local Lorentz invariance of GR has the role of a real gauge symmetry in the sense that the corresponding changes play no physical role in the space-time dynamics. Therefore, spin connections are not physical fields, but just gauge potentials, subject to such transformations that do not alter the curvature tensor, describing the Einsteinian meaning of the gravitational interaction. The diffeomorphism invariance of GR is related to the formal transcription of a physical property, i.e., the covariance of physical laws under changes of the reference system. Such an invariance is well formulated in terms of tensor quantities, vanishing or not in every reference frame. Spin connections transform like vectors, as far as diffeomorphisms are concerned, and can never vanish on a curved spacetime. Thus, if we are able to show, as we will argue in the following, that diffeomorphism invariance induces local Lorentz rotations, then we can conclude that spin connections can no longer be regarded to as good gauge potentials for these rotations, because of their transformation properties. In particular, on flat spacetime, these variables can be taken as vanishing, by choosing the tetrad vectors as $e^{a}_{\mu}=\delta^{a}_{\mu}$, such that they must remain identically zero under diffeomorphisms. For these coordinate transformations, which can be interpreted as local rotations, this behavior makes them unappropriate to restore local Lorentz invariance. In such a scheme, the request for external fields of Lorentz gauge connections appears well grounded. This picture remains still valid in curved spacetime, as long as we accept that the ambiguity of spin-connection transformations must be solved in favor of their tensor nature, when diffeomorphism-induced rotations are implemented. By other words, under real gauge transformations, spin connections feel a gauge transformation, but the external fields transform only according to their Lorentz indices. Nevertheless, when local rotations are induced by coordinate changes, the only fields able to restore the request of Lorentz invariance are the latter. In fact, the nature of gauge potentials is naturally lost by gravitational connections, behaving like tensors only. We are now going to demonstrate that, both in the infinitesimal and in the finite case, such correspondence between coordinate changes and local rotations takes place only if we deal with isometric diffeomorphisms. This request is naturally expected if we want to reproduce a Lorentz symmetry, and we stress that it always holds for this restricted class of diffeomorphisms on a Minkowski space (at the ground of the present paradigm).  
\paragraph{Finite gauge transformations}
Let us consider the coordinate transformation corresponding to the special case of an isometry of the spacetime, i.e, $x'=x'(x)$. The conditions ensuring invariance of the metric tensor in term of the bein vectors is
\begin{equation}
e^{a}_{\mu'}dx^{\mu'}=\Lambda ^{a}_{b}(x)e^{b}_{\nu}(x)dx^{\nu},
\end{equation}
where $\Lambda$ denotes a local rotation. This condition is easily restated in terms of the following gradients
\begin{equation}
\frac{dx^{\mu'}}{dx^{\nu}}=\Lambda ^{a}_{b}(x)e^{b}_{\nu}(x)e^{\mu'}_{a}(x').
\end{equation}
As far as isometric diffeomorphisms are concerned, we get the key relation
\begin{equation}
e^{a}_{\nu}(x)=e^{a}_{\mu'}\frac{dx^{\mu'}}{dx^{\nu}}=\Lambda ^{a}_{b}(x)e^{b}_{\nu}(x). 
\end{equation}
This way, we see that the isometric component of a diffeomorphism is formally indistinguishable from a local rotation of the bein basis, i.e., the physical implementation of a local Lorentz gauge.

\paragraph{Infinitesimal case}
The explanation of the puzzling scenario depicted at the end of Section (\ref{par1}) is hinted by the relation between diffeomorphisms and local Lorentz transformations. A generic diffeomorphism $\phi:M^4\rightarrow M^4$ maps the orthonormal basis fields $e_{\mu}^{\phantom1a}$ into $\phi^{*}\left(e_{\mu}^{\phantom1a}\right)$, which are not orthonormal in every point of the manifold, and do not represent any family of physically-realizable observers; on the contrary, an isometric diffeomorphism induces orthonormal transformed basis: an isometry generates a local Lorentz transformation of the basis. An infinitesimal isometric diffeomorphism, described by
\begin{equation}
x^{\prime\mu}=x^{\mu}+\xi^{\mu}(x),\qquad\text{and}\qquad\nabla_{\mu}\xi_{\nu}+\nabla_{\nu}\xi_{\mu}=0,
\end{equation}
induces a transformation of the basis vectors, 
\begin{equation}\label{taylor}
e^{\prime\mu}_{\phantom1\phantom2a}(x^{\prime})=e^{\nu}_{\phantom1a}(x)+e^{\nu}_{\phantom1a}(x)\partial_{\nu}\xi^{\mu}(x).
\end{equation}
Since
\begin{equation}
e^{\prime\mu}_{\phantom1\phantom2a}(x)=\Lambda_{a}^{\phantom1b}(x)e^{\mu}_{\phantom1b}(x)=\left(\delta^b_a +\epsilon^{b}_{\phantom1a}\right)e^{\mu}_{\phantom1b}(x),
\end{equation}
the rotation coefficients $\epsilon^{b}_{\phantom1a}$ of an infinitesimal Lorentz transformation can be written as a function of both the tetrad fields and the generic displacement $\xi^{\mu}(x)$.

On curved space-time, there exist two different local Lorentz transformations, which coincide in general flat space-time\footnote{It is worth remarking that active and passive translations are discussed within the framework of translation gauge too, as outlined, for example, in \cite{gag}.}.

Active Lorentz transformations are due to the action of the Lorentz group on vectors $V^{\mu}$ and spinors $\psi$ on the tangent bundle, i.e., $V^{\mu}\rightarrow \Lambda(x)^{\mu}_{\ \ \nu}V^{\nu}$ and $\psi\rightarrow S(\Lambda)(x)\psi$, respectively, and are mathematically represented by a Lorentz matrix depending on position and defined everywhere, 
\begin{align}
\nonumber W & ^{a_1a_2\dots}_{b_1b_2\dots}=\Lambda\left( x\right)^{a_1}_{\phantom1c_1}\dots\Lambda^{-1}\left( x\right)_{b_1}^{\phantom1d_1}\dots W^{c_1c_2\dots}_{d_1d_2\dots}\,,
\\[8pt]
&\psi^{\prime}=S\left(\Lambda\left( x\right)\right)\psi, \qquad \overline\psi^{\prime}=\overline\psi S^{-1}\left(\Lambda\left(x\right)\right).
\end{align}
Passive Lorentz transformations are due to isometric diffeomorphisms of the manifold, which pull-back the local basis in the generic point $P$, 
\begin{equation}
\phi_{*}\left(e^a\right)=\Lambda\left( x\right)^{a}_{\phantom1b}e^b\Rightarrow \Lambda\left(P\right)^{a}_{\phantom1b}=e_{\mu}^{\phantom1a}\left.\frac{\partial x^{\mu}}{\partial x^{\prime\nu}}\right|_{x^{\prime}=P}e_{\phantom1b}^{\nu}\,.
\end{equation}
While active transformations do not involve coordinates and are defined everywhere once the matrix function $\Lambda\left(x\right)^{a}_{\phantom1b}$ is assigned, passive transformations can be reduced to a local Lorentz transformation only in the generic point $P$, acting as a pure diffeomorphisms in the other points on the manifold.

Active and passive Lorentz transformations can be demonstrated to coincide in curved space-time too, as it can be inferred by the comparison of the transformation laws for the tetrad field under Lorentz and world transformations. In fact, for local Lorentz transformations, one gets
\begin{equation}
e'^{a}_{\mu}(x')= \Lambda^{a}_{\ b}(x')e^{a}_{\mu}(x'),
\end{equation}
while, for world transformations,
\begin{equation}
e^{a}_{\mu}(x)\rightarrow e'^{a}_{\mu}(x')=e^{a}_{\mu}(x)\frac{\partial x^{\rho}}{\partial x'^{\mu}}\approx e^{a}_{\mu}(x)+e^{a}_{\rho}(x)\frac{\partial \xi^{\rho}}{\partial x'^{\mu}}.
\end{equation}
The comparison leads to
\begin{equation}
e'^{a}_{\mu}(x')=
e^{a}_{\mu}(x')+e^{b}_{\mu}(x')\epsilon^{a}_{b},
\end{equation}
where
\begin{equation}\label{lkjh}
\epsilon^{a}_{\ b}\equiv  -D_{b}\xi^{a}-R^{a}_{\ bc}\xi^{c}
\end{equation}
and $\lambda_{abc}=R_{abc}-R_{bac}$ are the anholonomy coefficients. To pick up local Lorentz transformations from the set of generic diffeomorphisms, the isometry condition $\nabla_{(\mu}\xi_{\nu)}=0$ must be taken into account, so that in (\ref{lkjh}) only the anti-symmetric part of $D_{b}\xi^{a}$ does not vanish. Finally, we get
\begin{equation}\label{freddy}
\epsilon_{ab}=D_{[a}\xi_{b]}-R_{abc}\xi^{c},
\end{equation}
which is anti-symmetric, i.e. $\epsilon_{ab}=-\epsilon_{ba}$.

\paragraph{Discussion}
These considerations helped us focus general attention on the use of local passive Lorentz rotations, and search a consistent formulation for the physical nature of the corresponding gauge fields. The difference between flat and curved spacetime is emphasized because, in the former case, we are allowed to extract physical information on the proposed gauge fields in the simple case of vanishing spin connections, while, in the second case, details on the interaction between gravity and gauge fields can be outlined by fixing the extended dynamical equations.

\section{Gauge theory of the Lorentz group on flat Space-Time}\label{par2}

This section is dedicated to the construction of a gauge description of the pure Lorentz group on a flat Minkowski space-time.
The choice of flat space is due to the fact that, in this case, the Riemann curvature tensor vanishes and, consequently, the usual spin connections $\omega^{a b}$ can be set to zero (but, in general, they are allowed to be non-vanishing in view of local Lorentz invariance). This allows one to introduce Lorentz-valued connections as the gauge field of passive local Lorentz transformations on flat space-time as far as the correspondence between an infinitesimal diffeomorphism and a local Local rotation is recovered, as shown in the previous section.

\paragraph{Tensors}\null
Let $M^{4}$ be a 4-dimensional flat manifold: the metric tensor $g_{\mu\nu}$ reads
\begin{equation}\label{gen}
g_{\mu\nu}=\eta_{\alpha\beta}\frac{\partial x^{\alpha}}{\partial y^{\mu}}\frac{\partial x^{\beta}}{\partial y^{\nu}}=\eta_{\alpha\beta}e^{\alpha}_{\mu}e^{\beta}_{\nu},
\end{equation}
where $e^{\alpha}_{\mu}$ are bein vectors, $x^{\alpha}$ are Minkowskian coordinates, and $y^{\mu}$ are generalized coordinates. For an infinitesimal generic diffeomorphism
\begin{equation}\label{gendifff}
y^\mu\equiv x^\alpha\rightarrow y'^{\mu}=y'^\mu(x^\alpha)=x^{\alpha} +\xi^{\alpha}(x^{\gamma})
\end{equation}
and for an infinitesimal local Lorentz transformation
\begin{equation}
x^{\alpha}\rightarrow x'^{\alpha}=x^{\alpha} +\epsilon^{\alpha}_{\ \beta}(x^{\gamma})x^{\beta},
\end{equation}
the behavior of a vector field $V_{\alpha}\rightarrow V'_{\alpha}$ must be the same:
from the comparison of the two transformation laws 
\begin{equation}
V'_{\mu}(y'^{\rho})= V_{\mu}(y^{\rho})+\frac{\partial \xi^{\nu}(y^{\rho})}{\partial y^{\mu}}V_{\nu}(y^{\rho}),
\end{equation}
\begin{equation}
V'_{\alpha}(x'^{\gamma})= V_{\alpha}(x^{\gamma})+\epsilon^{\ \beta}_{\alpha}V_{\beta}(x^{\gamma}),
\end{equation}
respectively, the identification 
\begin{equation}\label{loc}
\epsilon _{\alpha}^{\ \beta}\equiv \frac{\partial \xi_{\alpha}(x^{\gamma})}{\partial x^{\beta}}
\end{equation}
is possible, where the isometry condition
\small 
\begin{equation}
\partial_{\beta}\xi_{\alpha}+\partial_{\alpha}\xi_{\beta}=0
\end{equation}
\normalsize
has been taken into account in order to restore the proper number of degrees of freedom of Lorentz transformations, $10$, out of that of generic diffeomorphisms, $16$.

The coordinate transformation that induces vanishing Christoffel symbols in the point $P$ is
\begin{equation}\label{cris}
x^{\alpha}_{P}=x^{\alpha}_{tb}+\frac{1}{2}\left[\Gamma^{\alpha}_{\beta\delta}\right]_{P}x^{\beta}_{tb}x^{\delta}_{tb},
\end{equation}
where $tb$ refers to the tangent bundle:
the comparison with a generic diffeomorphism (\ref{gendifff})
leads to the identification in the point $P$
\begin{equation}\label{map}
y^\mu(x^{\alpha})_{P}=x^{\alpha}_{tb}+\frac{1}{2}\left[\Gamma^{\alpha}_{\beta\delta}\right]_{P}x^{\beta}_{tb}x^{\delta}_{tb}-\xi^{\alpha},
\end{equation}
i.e. the coordinates of the tangent bundle are linked point by point to those of the Minkowskian space through the relation (\ref{cris}), and they differ for the presence of the infinitesimal displacement $\xi$. 

\paragraph{Spinors} The action describing the dynamics of spin-$\frac{1}{2}$ fields on a 4-dimensional flat Minkowski manifold $M^{4}$ is 
\begin{equation}\label{Dirac's Lagrangian}
S=\frac{i}{2}\int d^4x\left(\overline{\psi}\gamma^{a}\partial_{a}\psi-(\partial_{a}\overline{\psi})\gamma^{a}\psi\right),
\end{equation}
where $\gamma^{a}$ are Dirac matrices, and is invariant under global Lorentz transformations, which act on the spinor fields $\psi$ and $\overline{\psi}$ as 
\begin{equation}\label{spinor transformation}
\psi\rightarrow S\left(\Lambda\right)\psi,\qquad\overline{\psi}\rightarrow
\overline{\psi}S^{-1}\left(\Lambda\right),
\end{equation}
where $S\left(\Lambda\right)$ is a nonsingular function of the Lorentz matrix $\Lambda$, if the $\gamma$ matrices transform like vectors,
\begin{equation}\label{gamma vector}
S\left(\Lambda\right)\gamma^{a}S^{-1}\left(\Lambda\right)=\Lambda^{a}_{\phantom1b}\gamma^{b}.
\end{equation}

$S\left(\Lambda\right)$ for infinitesimal proper transformations ($\left(\Lambda\right)^{a}_{\phantom1b}=\delta^{a}_{\phantom1b}+\epsilon^{a}_{\phantom1b}$, with $\epsilon^{ab}=-\epsilon^{ba}$ and $\epsilon^{a}_{\phantom1b}\ll1$) reads
\begin{equation}
S\left(\Lambda\right)=1-\frac{i}{4}\epsilon^{ab}\Sigma_{ab},
\end{equation}
where $\Sigma_{ab}$ are the generators of the Lorentz group\footnote{In analogy with the formalisms of particle Physics and renormalization techniques \cite{mandl,weinberg}, a suitable coupling constant could be attributed to the symmetry induced by the Lorentz group. Anyhow, because of the technical character of this analysis, here we prefer follow the notation of the great majority of the works \cite{wal84,shap}, also in metric-affine gravity. Nonetheless, it is worth remembering that such a coupling constant should be very small, as this kind of interaction has not been detected experimentally yet \cite{oba83}. For some issues related to the use of such a coupling constant, see also \cite{oba83}.}.

If this scenario is generalized, i.e., if accelerated coordinates, which are related to Minkowskian coordinates through (\ref{gen}), are taken into account, spinor fields, differently from vector fields, have to recognize the isometric components of the diffeomorphism (\ref{loc}) as a local Lorentz transformation. In fact, bein vectors $e^{\alpha}_{\mu}$ defined in (\ref{gen}) form a basis on $M^{4}$, and, as a vector field, are the projection operators that map each point of the manifold to the tangent bundle, as defined in (\ref{map}), thus characterizing the space where local Lorentz transformations live.\footnote{It is worth remarking that, from an operationally-motivated point of view, bein vectors characterize a family of observers; in particular, the vector $e^{0}_{\mu}$, at each event, is tangent to the world line of the observer at that point, while the 3-dimensional basis vectors $e^{i}_{\mu}$ ($i\equiv1,2,3$) are aligned along the principal axis of the experimental device. In the accelerated frame, the diffeomorphism-induced Lorentz group is local, i.e., at each point of the manifold, there is a different action of the Lorentz group: the equivalence of every local basis ($\equiv$observer) introduces on the flat manifold $M^{4}$ a gauge freedom connected with pure local Lorentz rotations.} 
In an accelerated frame, Lorentz-valued connections have to be introduced for matter fields: 
\begin{equation}\label{Dirac in curv}
L=\frac{i}{2}\,e^{\mu}_{\phantom1a}\left[\overline{\psi}\gamma^{a}\partial_{\mu}\psi-\partial_{\mu}\overline{\psi}\gamma^a\psi\right],
\end{equation}
where the projectors $e^{\mu}_{\phantom1a}$ from the target space to the tangent physical space are present; local Lorentz transformations for spinor fields read
\begin{equation}\label{spinor local transform}
\psi(x)\rightarrow S(\Lambda)(x)\psi(x),\quad\overline\psi(x)\rightarrow
\overline\psi(x)S^{-1}\left(\Lambda\right)(x),
\end{equation}
where $S(\Lambda)(x)$ is a non-singular matrix $\forall x$. Let us assume that the $\gamma$ matrices transform locally as vectors, i.e.,
\begin{equation}\label{gamma local transform}
S(\Lambda)(x)\gamma^{a} S^{-1}(\Lambda)(x)=\left(\Lambda^{-1}\right)^{a}_{\phantom1b}(x)\gamma^{b}:
\end{equation}
for an infinitesimal local Lorentz transformation
\begin{equation}
S\left(\Lambda\right)=1-\frac{i}{4}\epsilon^{ab}(x)\Sigma_{ab},
\end{equation}
covariant gauge derivatives
\begin{subequations}\label{cov dev}
\begin{align}
D_{a}\psi & =e^{\mu}_{\phantom1a}D_{\mu}\psi=e^{\mu}_{\phantom1a}\left(\partial_{\mu}\psi-\frac{i}{4}A_{\mu}^{b c}\Sigma_{b c}\psi\right),
\\
\overline{D_{a}\psi} & =e^{\mu}_{\phantom1a}\overline{D_{\mu}\psi}=e^{\mu}_{\phantom1a}\left(\partial_{\mu}\overline{\psi}+\frac{i}{4}\overline{\psi}\Sigma_{b c}A_{\mu}^{b c}\right),
\end{align}
\end{subequations}
assure invariance under local Lorentz transformations, 
\begin{equation}
\gamma^{\mu}D_{\mu}\Psi \rightarrow S(\Lambda)\gamma^{\mu}D_{\mu}\Psi
\end{equation}

provided that the gauge fields transform as

\begin{equation}
A^{a}_{\phantom1b}\rightarrow \Lambda(x)^{a}_{\phantom1c}A^{c}_{\phantom1d}\Lambda^{-1}(x)^{d}_{\phantom1b}+\Lambda(x)^{a}_{\phantom1c}d\Lambda^{-1}(x)^{c}_{\phantom1b},
\end{equation}
or, in the infinitesimal case,
\begin{equation}
\omega_{\mu}\rightarrow S^{-1}(\Lambda)\omega_{\mu}S^(\Lambda)-S^{-1}(\Lambda)\partial_{\mu}S(\Lambda),
\end{equation}
i.e., as Yang-Mills gauge fields.

The implementation of local Lorentz symmetry, $\partial_{\mu}\rightarrow D_{\mu}$, leads to the interaction-Lagrangian density
\begin{equation}
\mathcal{L}_{int}=\frac{1}{8}\left(\overline{\psi}\gamma^{a} \Sigma_{bc}A^{bc}_{\mu}e^{\mu} _{\phantom1a}\psi-\overline{\psi} \Sigma_{bc}\gamma^{a}A^{bc}_{\mu}e^{\mu}_{\phantom1a}\psi\right),
\end{equation}
which can be equivalently rewritten as
\begin{equation}
\mathcal{L}_{int}=\frac{1}{8}\,e^{\mu} _{\phantom1a}\,\overline{\psi}\left\{\gamma^{a},\Sigma_{b c}\right\}\psi A^{b c}_{\mu}=-J_{b c}^{\mu}A^{b c}_{\mu},
\end{equation}
where curl brackets $\{\}$ indicate the anti-commutator, and, because
\begin{equation}\label{anti commutator}
\left\{\gamma^a,\Sigma^{b c}\right\}=2\epsilon^{a b c}_{\phantom1\phantom1\phantom1d}\gamma_5\gamma^d,
\end{equation}
we get 
\begin{equation}
J^{a b}_{\mu}=-\frac{1}{4}\,\epsilon^{a b}_{\phantom1\phantom1c d}e_{\mu}^{\phantom1c}j_{A}^d,
\end{equation}
where $j_{A}^d=\overline{\psi}\gamma_5\gamma^d\psi$ is the spinor axial current: the spinor axial current interacts with the gauge field $A_{\mu}$, and is the source of the gauge field itself.

An action for the gauge field has to be added: since the curvature 2-form of the Lorentz gauge connections 
\begin{equation}
F^{ab}=d A^{ab}+A^{a}_{\phantom1c}\wedge A^{c b},
\end{equation}
is not invariant under gauge transformations, as usual in Yang-Mills gauge theories,
the gauge invariant action for the model will be
\begin{equation}\label{action for A}
S\left(A\right)=\frac{1}{32}\int  tr\,\star F\wedge F=-\frac{1}{4}\int d^4x\,\det\left\{e\right\}F_{\mu\nu}^{\phantom1\phantom2ab}F^{\mu\nu}_{\phantom1\phantom2ab},
\end{equation}
where $\star$ denotes the Hodge operator. From a physical point of view, the most natural action is (\ref{action for A}) \cite{weinberg}. Anyhow, from a mathematical and more abstract point of view, it would also be possible to introduce the irreducible pieces of $F$ with different weights. We will further discuss this possibility within the framework of curved spacetime.

\paragraph{Field equations}\null
Collecting all the terms together, we obtain the complete action 
\begin{align}
S\left(A\right)+S_{FM}\left(\psi,\overline{\psi},A\right)
=\int d^{4}x\, det\left\{e\right\}\left\{-\tfrac{1}{4}\,\left(F_{\mu\nu}^{\phantom1\phantom2a b}F^{\mu\nu}_{\phantom1\phantom2a b}\right)
+\tfrac{i}{2}\,e^{\mu}_{\phantom1a}\left[\overline\psi\,\gamma^{a}D_{\mu}\psi-D_{\mu}\overline{\psi}\,\gamma^{a}\psi\right]\right\},
\end{align}
where the covariant derivatives (\ref{cov dev}) have been introduced. It is straightforward to verify that this expression naturally fits all the features of a Yang-Mills gauge description. In fact, the covariant derivatives (\ref{cov dev}) assure invariance under local Lorentz transformations, in terms of a gauge transformation, for the spinor part of the action, and the term (\ref{action for A}) also is invariant under such transformations. According to this picture, it will be natural to obtain the typical field equations of a Yang-Mills theory. Furthermore, it is worth remarking that the introduction of different irreducible pieces of $F$ with different weights would spoil such gauge description.

Since we are dealing, for the moment, with flat space-time, tetrad vectors are not dynamical fields, but only projectors from the target space to the general physical space, then they will appear only in the expression of the invariant volume of the space-time and in scalar products: no variation with respect to them will be needed for field equations. Actually the only real dynamical field is the Lorentz-valued connection 1-forms $A^a_{\phantom1b}$. In fact, if, in analogy with GR, the curvature 2-form saturated on bein vectors is considered as an action for the model, a trivial theory is obtained. Variation of (\ref{demu}) with respect to the tetrad field would provide the total energy-momentum tensor accounting for the dynamics and interactions of the vector field $A$ and the spinor field $\Psi$, respectively. Such a variation will indeed be crucial in casting Einstein equations on curved space-time (where the gravitational action has to be added), but here the vier-bein variables are regarded simply as ' kinematic ' variables only.

Variation with respect the field $A_{\mu\phantom1b}^{\phantom1a}$ leads to the dynamical equations
\begin{equation}\label{demu}
D_{\mu}F^{\mu\nu a}_{\phantom1\phantom1\phantom1\phantom1b}=J^{\nu a}_{\phantom1\phantom1b},
\end{equation}
which are the Yang-Mills equations for the non-Abelian gauge field of the Lorentz group on flat space-time. The source of this gauge field is the conserved density of spin of the fermion matter.

Variation with respect to the spinor fields $\psi$ and $\overline\psi$ and relation (\ref{anti commutator}) lead to the usual Dirac interaction equations for the spinor field and for the adjoint field:
\begin{equation}
e^{\mu}_{\phantom1a}\left[i\gamma^{a}\partial_{\mu}+\frac{1}{8}\left\{\gamma^a,\Sigma_{c d}\right\}A^{c d}_{\mu}\right]\psi=e^{\mu}_{\phantom1a}\left[i\gamma^{a}\partial_{\mu}+\frac{1}{4}\,\epsilon^{a b}_{\phantom1\phantom1c d}\gamma_5\gamma_b A^{c d}_{\mu}\right]\psi=0,
\end{equation}
and
\begin{equation}
e^{\mu}_{\phantom1a}\overline{\psi}\left[i\gamma^{a}\stackrel{\leftarrow}{\partial}_{\mu}-\frac{1}{8}\left\{\gamma^a,\Sigma_{c d}\right\}A^{c d}_{\mu}\right]=e^{\mu}_{\phantom1a}\overline{\psi}\left[i\gamma^{a}\stackrel{\leftarrow}{\partial}_{\mu}-\frac{1}{4}\,\epsilon^{a b}_{\phantom1\phantom1c d}\gamma_5\gamma_b A^{c d}_{\mu}\right]=0.
\end{equation}

Field equations illustrate that the dynamics for a spinor field in an accelerated frame differs from the standard Dirac dynamics for the spinor-gauge field interaction term, i.e., spinor fields are not free fields any more. For the analysis of the Dirac equation in non-inertial systems in flat spacetime, see also \cite{lus2}. 
The present goal is to extend this formulation on a curved space-time manifold, on which non-vanishing Ricci rotation coefficients appear.

\section{Gauge theory of the Lorentz group on curved Space-Time}\label{par3}
The considerations developed in the previous sections can be generalized to curved space-time; the torsion-less assumption of GR, perfectly realized by the Hilbert-Palatini action, does not allow for an independent gauge field of the Lorentz group.

The connections $A^a_{\phantom1b}$ have been introduced on a general flat manifold in order to restore local invariance for the Lagrangian density of spinor fields under passive local Lorentz transformations. Generalizing the framework to curved space-time will provide a geometrical interpretation for the new connection fields, which will be identified with the field $K^a_{\phantom1b}$. The need to introduce local Lorentz gauge fields in curved space-time is aimed at restoring local invariance of the spinor Lagrangian density under passive local Lorentz transformations, while spin connections allow one to recover the proper Dirac algebra for Dirac matrices. The generalization consists in considering the space-time $M^4$ as a curved manifold, on which the tetrad basis is consists of dynamical fields, which describe pure gravity; local Lorentz transformations are still considered as a gauge freedom, so that Lorentz-valued connection fields have to be introduced.

In the next paragraphs, within the framework of curved space-time, the relation between the gauge field of the Lorentz group and the geometrical properties of metric-compatible space-times will be investigated. In particular, in the second-order approach, the possibility of identifying the contortion field with Lorentz connections will be investigated, while, in the first-order approach, the geometrical hypotheses for the introduction of torsion as a Lorentz gauge field  will be addressed. The two approaches will be compared in the linearized regime.

  
\paragraph{Second-order approach}\null
If we consider Riemann-Cartan spaces, endowed with the affine connections (\ref{cane}),
we look for an operator $D_{\mu}$ which allows for
\begin{equation}
D_{\mu}\gamma_{\nu}=0;
\end{equation}
such an operator is found to be
\begin{equation}\label{gatto}
D_{\mu}A=\nabla_{\mu}A-[\Gamma_{\mu},A]
\end{equation}
for a generic geometrical object, and
\begin{equation}\label{canarino}
D_{\mu}\psi=\partial_{\mu}\psi-\Gamma_{\mu}\psi,\quad D_{\mu}\bar{\psi}=\partial_{\mu}\bar{\psi}+\bar{\psi}\Gamma^{\mu}
\end{equation}
for spinor fields, so that the matter Lagrangian density reads
\begin{equation}
L_{M}=-\frac{i}{2}\bar{\psi}\gamma^{a}e^{\mu}_{a}D_{\mu}\psi +H.C.,
\end{equation}
where $H.C.$ denotes Hermitian conjugation.
 
By substitution of (\ref{cane}) in (\ref{gatto}), after standard manipulation one finds the expression for the connections
\begin{equation}
\Gamma_{\mu}=\Gamma_{\mu}^{R}+\Gamma_{\mu}^{K}=
\frac{1}{2}R^{a\bar{b}}_{\mu}\Sigma_{ab}+\frac{1}{2}A^{a\bar{b}}_{\mu}\Sigma_{ab},
\end{equation}
where the tetrad projection of the Ricci coefficients and of the contortion field are defined, respectively,
\begin{equation}
R_{ab\mu}=R_{abc}e^{c}_{\mu},
\end{equation}
\begin{equation}\label{cappa}
A_{ab\mu}\equiv -K_{\rho\sigma\mu}e^{\rho}_{a}e^{\sigma}_{b}.
\end{equation}
The connections $\Gamma_{\mu}$ defined in (\ref{gatto}) split up into two different terms, the spin connections $\Gamma_{\mu}^{R}$, which restore the commutation relations of the Dirac matrices in the physical space-time, and the gauge connections $\Gamma_{\mu}^{K}$, which reestablish invariance under local Lorentz transformations, respectively. If $R=0$, the scenario depicted above reduces to the results of (\ref{par2}), so that the gauge connections $\Gamma_{\mu}^{K}$ can be interpreted as the real gauge fields of the local Lorentz group, for they are non-vanishing quantities even in flat space-time, as requested for any gauge field. Furthermore, formula (\ref{freddy}) illustrates once more that $R_{abc}$ cannot be a gauge field, for it defines gauge transformations on the tangent bundle.

Since,in a gauge setting, gauge connections are primitive objects, the total action $S \equiv S(e,A,\psi)$ must depend on the independent fields $\psi$, $e$, and $A$, such as
\begin{equation}\label{svaria}
S  = {S}(e,A,\psi) = -\frac{1}{2}\int det(e)d^{4}x[R(e)- (L_{M}-\frac{1}{4}F_{\mu\nu}(A)F^{\mu\nu}(A))],
\end{equation}
where gauge-fixing terms have not been included, as it will be convenient, for our purposes, to work in the restricted space of conserved currents.
 Variation of the action with respect to the independent fields leads to field equations.
Variation with respect to bein vectors, $\delta e^{a}_{\mu}$, leads to the bein projection of the Einstein equations, with Yang-Mills tensor $T^{\mu\nu}$  as source
\begin{equation}
(R_{\mu\nu}-\frac{1}{2}g_{\mu\nu}R)e^{\nu}_{a}=T_{\mu a},
\end{equation}
while variation with respect to the field $\delta A_{\mu}^{ab}$ brings Yang-Mills equations, with the spinor current density as a source:
\begin{equation}\label{bo}
D_{\mu}F^{\mu\nu a}_{\phantom1\phantom1\phantom1\phantom1b}=J^{\nu a}_{\phantom1\phantom1b}.
\end{equation}
Finally, the Dirac equation for the spinor field $\psi$ is obtained after variation with respect to the adjoint field, $\delta \bar{\psi}$,
\begin{equation}
\gamma^{\mu}D_{\mu}\psi=0,
\end{equation}
and vice-versa.

The comparison between local Lorentz transformations and gauge transformations allows one to obtain the expression for conserved quantities. This way, since the current density (\ref{bo}) admits the conservation law
\begin{equation}
D_{\mu}J^{\mu ab}=0,
\end{equation}
a conserved (gauge) charge\footnote{This quantity is a conserved one if one assumes that the fluxes through the boundaries of the space integration vanish} can be defined
\begin{equation}\label{mah}
Q^{ab}=\int d^{3}x J^{0 ab}=const;
\end{equation}
on the other hand, the bein projection of the spin term of the angular momentum tensor $M^{\mu\nu}$, the conserved quantity for Lorentz transformations in flat space time, reads
\begin{equation}\label{mahh}
M^{ab}= \int  d^{3}x  \pi_{r}\Sigma^{ab}_{rs}\psi_{s}= const.,
\end{equation}
which coincides with (\ref{mah}), provided that $\pi_{r}$ is the  density of momentum conjugate to the field $\psi_{r}$, i.e., $\pi_{r}=\partial L / \partial \dot{\psi}_{r}$. This identification is possible only on flat space-time, because of the definition of the parameter $\epsilon^{ab}$ (\ref{freddy}), which points up the remarkable features of local Lorentz transformations on the tangent bundle.

\paragraph{First-order approach}\null
If one relaxes the torsion-less assumption, the second Cartan structure rewrites
\begin{equation}\label{Cartan eq general}
d e^{a}+\omega^{a}_{\phantom1b}\wedge e^{b}=T^{a},
\end{equation}
where $T^{a}$ is the torsion 2-form; this equation is solved by the connections
\begin{equation}\label{connection}
\omega^a_{\phantom1b}=\widetilde{\omega}^a_{\phantom1b}+K^a_{\phantom1b},
\end{equation} 
where $K^{a}_{\phantom1b}$ is the contortion 1-form, such that $T^{a}=K^{a}_{\phantom1b}\wedge e^{b}$, while $\widetilde{\omega}^a_{\phantom1b}$ are the usual connection 1-forms. As a result, new 1-forms appear in the dynamics, which reestablish the proper degrees of freedom for the connections of the Lorentz group.

In GR, nevertheless, these connections do not describe any physical field: after substituting the solution (\ref{connection}) of the structure equation into the Hilbert-Palatini action\footnote{Let $S\left(q_i,Q_j\right)$ be an action depending on two sets of dynamical variables, $q_i$ and $Q_j$. The solutions of the dynamical equations are extrema of the action with respect to both the two sets of variables: if the dynamical equations $\partial S/\partial q_i=0$ have a unique solution, $q_i^{(0)}\left(Q_j\right)$ for each choice of $Q_j$, then the extrema of the pullback $S\left(q_i\left(Q_j\right),Q_j\right)$ of the action to the set of solution are precisely the extrema of the total total action $S\left(q_i,Q_j\right)$. For an application of this theorem, see, for example, \cite{ash89}.}, one finds that the connections $K^a_{\phantom1b}$ appear only in a non-dynamical term, i.e.,
\begin{equation}
S(e,\omega)=\frac{1}{2}\int\epsilon_{a b c d}\,e^{a}\wedge e^{b}\wedge\left(\widetilde{R}^{c d}+K^{c}_{f}\wedge K^{f d}\right).
\end{equation}
Since the other terms vanish because of the structure equation, the connections $K^a_{\phantom1b}$ themselves vanish after variation, unless spinors are taken into account: in this case, the connections $K^a_{\phantom1b}$ become proportional to the spin density of the matter, thus giving rise to the Einstein-Cartan model, where the usual four-fermion term arises.

As far as the formulation of a Lorentz gauge theory is concerned, we will denote the connection 1-forms for local Lorentz transformations, independent of any other fields, with the quantities $A^a_{\phantom1b}$ , and the standard connections of GR, which depend on the gravitational field and on spinors (if matter is taken into account) with $\omega^a_{\phantom1b}$. Then the total connections eventually rewrite
\begin{equation}\label{connection-1}
C^a_{\phantom1b}=\omega^a_{\phantom1b}+A^a_{\phantom1b}.
\end{equation}

From the comparison of (\ref{connection}) and (\ref{connection-1}), it can be inferred that the presence of the fields $A$ is connected with the appearance of torsion in space-time, so that, if the proper geometrical interpretation has to be attributed to the fields $A$, the interaction term between the spin connections $\omega$ and the fields $A$
\begin{equation}\label{interacting term}
S_{int}=2\int\epsilon_{a b c d}\,e^{a}\wedge e^{b}\wedge\omega^{[c}_{\phantom1f}\wedge A^{f d]}
\end{equation}
has to be postulated.
The action describing the dynamics of the fields $A^a_{\phantom1b}$ is just the same introduced in section (\ref{par2}), (\ref{action for A}), while the action that accounts for the connections $\omega^a_{\phantom1b}$ can be taken as (\ref{action for o}), since its presence is due to the existence on curved space-time of a particular local Lorentz transformation connected to the invariance under diffeomorphisms. Collecting all the terms together, we get
\begin{align}\label{total action}
\nonumber 
& S\left(e,\omega,A,\psi,\overline{\psi}\right)=  \frac{1}{4}\int\epsilon_{a b c d}\,e^{a}\wedge e^{b}\wedge R^{c d}+\\
& -\frac{1}{32}\int tr\,\star F\wedge F-\frac{1}{4}\int\epsilon_{a b c d}\,e^{a}\wedge e^{b}\wedge\omega^{[c}_{\phantom1f}\wedge A^{f d]}+
\nonumber \\
& +\frac{1}{2}\int\epsilon_{a b c d}\,e^{a}\wedge e^{b}\wedge e^{c}\wedge\left[i\overline{\psi}\gamma^d\left(d-\frac{i}{4}\left(\omega+A\right)\right)\psi-i\left(d+\frac{i}{4}\left(\omega+A\right)\right)\overline{\psi}\gamma^d\psi\right].
\end{align}
Two cases can be distinguished, where the properties of the Lorentz connections are defined from a geometrical point of view, according  to the absence or presence of spinors.

If fermion matter is absent, the action reduces to 
\begin{equation}\label{action vacuum}
S\left(e,\omega,A\right)=\frac{1}{2}\int\epsilon_{a b c d}\,e^{a}\wedge e^{b}\wedge R^{c d}-\frac{1}{32}\int tr\,\star F\wedge F-\int\epsilon_{a b c d}\,e^{a}\wedge e^{b}\wedge\omega^{[c}_{\phantom1f}\wedge A^{f d]}.
\end{equation}
Variation with respect to the connections $\omega$ gives, after standard calculations,
\begin{equation}\label{equation for omega}
d^{(\omega)}e^a=A^a_{\phantom1b}\wedge e^b,
\end{equation}
which admits the solution
\begin{equation}\label{solution vacuum}
\omega^a_{\phantom1b}=\widetilde{\omega}^a_{\phantom1b}+A^a_{\phantom1b},
\end{equation}
were $\widetilde{\omega}^a_{\phantom1b}$ are the usual connection 1-forms: because of the analogy with the solution of the second Cartan structure equation (\ref{connection}), the connection  $A$ can be identified with the 1-form $K$. 

Since solution (\ref{solution vacuum}) is unique, action (\ref{action vacuum}) can be pulled back to the given solution to obtain the reduced action for the system, which now depends on the gravitational field and on the independent connections of the Lorentz group only. Namely, we have:
\begin{align}\label{reduced action}
\nonumber & S\left(e,A\right)=\frac{1}{4}\int\epsilon_{a b c d}\,e^{a}\wedge e^{b}\wedge\widetilde{R}^{c d}-\frac{1}{32}\int tr\,\star F\wedge F+\\
&-\frac{1}{4}\int\epsilon_{a b c d}\,e^{a}\wedge e^{b}\wedge\widetilde{\omega}^{[c}_{\phantom1f}\wedge A^{f d]}-\frac{1}{4}\int\epsilon_{a b c d}\,e^{a}\wedge e^{b}\wedge A^{c}_{\phantom1f}\wedge A^{f d},
\end{align}
where $\widetilde{\omega}=\widetilde{\omega}(e)$ and $\widetilde{R}=\widetilde{R}(\widetilde{\omega})$ denote the Ricci spin connections and the Riemann curvature 2-form, respectively.

Variation with respect to the gravitational field and the connections of the Lorentz group leads to
\begin{subequations}\label{dynamical equations}
\begin{align}\label{Einstein total equations}
\epsilon^a_{\phantom1b c d}\,e^{b}\wedge\widetilde{R}^{c d} & =M^a+\epsilon^a_{\phantom1b c d}\,e^{b}\wedge\left(\widetilde{\omega}^{c}_{\phantom1f}+A^{c}_{\phantom1f}\right)\wedge A^{f d},
\\
\label{ymabs}d^{(A)}\star F^{f d} & =\epsilon_{a b c}^{\phantom1\phantom1\phantom1[d}\,e^a\wedge e^b\wedge\left(\omega^{c f]}+2A^{c f]}\right),
\end{align}
\end{subequations}
where $M^a$ is the energy-momentum 3-form of the field $A$, which can be explicitly obtained after variation of the Yang-Mills-like action with respect the gravitational 1-form.
Eq.s (\ref{dynamical equations}) describe the coupled system of gravitational- and Lorentz-connection fields, i.e., they do not couple only through the energy-momentum tensor of the connection field. In fact, the presence of the interaction term (\ref{interacting term}) yields non-standard couplings both on the rhs of the Einstein equations and in the rhs of the Yang-Mills dynamical equations, and, in particular, while in flat space, the only source for the Lorentz connection fields $A$ is the density of spinor matter, in curved space-time also the gravitational spin connections become a source for this field. As a result, the gravitational field is a source for the torsion of space-time.

When the matter contribution is taken into account, variation of (\ref{total action}) with respect to the connections $\omega$ leads to
\begin{equation}\label{general equation for omega}
d^{(\omega)}e^a=A^a_{\phantom1b}\wedge e^b-\frac{1}{4}\epsilon^a_{\phantom1b c d}e^b\wedge e^c j_{(A)}^d,
\end{equation}
where $j_{(A)}^a=\overline{\psi}\gamma_5\gamma^a\psi$, i.e., the spinor axial current, deeply modifies eq. (\ref{equation for omega}). In fact, the presence of spinor matter prevents one from identifying the connections $A$ as the only generator of torsion, since all the terms in the rhs of the second Cartan structure equation have to be interpreted as torsion. This way, both the fields $A$ and the spinor axial current contribute to the torsion of space-time. It is worth noting that, if the field $A$ vanishes, we obtain the usual result of PGT, i.e. the Einstein-Cartan contact theory, in which torsion is directly connected with the density of spin and does not propagate.

Eq. (\ref{general equation for omega}) admits the unique solution
\begin{equation}
\omega^a_{\phantom1b}=\widetilde{\omega}^a_{\phantom1b}+A^a_{\phantom1b}+\frac{1}{4}\epsilon^a_{\phantom1b c d}e^c j_{(A)}^d,
\end{equation}
which can be inserted in the total action (\ref{total action}), thus bringing the result
\begin{align}\label{total action reduced}
\nonumber & S\left(e,A,\psi,\overline{\psi}\right) =\frac{1}{4}\int\epsilon_{a b c d}\,e^{a}\wedge e^{b}\wedge \widetilde{R}^{c d}-\frac{1}{32}\int tr\,\star F\wedge F+
\\\nonumber
& +\frac{1}{2}\int\epsilon_{a b c d}\,e^{a}\wedge e^{b}\wedge e^{c}\wedge\left[i\overline{\psi}\gamma^d\left(d-\frac{i}{4}\left(\widetilde{\omega}+A\right)\right)\psi-i\left(d+\frac{i}{4}\left(\omega+A\right)\right)\overline{\psi}\gamma^d\psi\right]
\\\nonumber
& -\frac{1}{4}\int\epsilon_{a b c d}\,e^{a}\wedge e^{b}\wedge A^{c}_{\phantom1f}\wedge A^{f d}-\frac{1}{4}\int\epsilon_{a b c d}\,e^{a}\wedge e^{b}\wedge\widetilde{\omega}^{[c}_{\phantom1f}\wedge A^{f d]}+
\\
& -\frac{3}{16}\int e_a\wedge e_b\wedge e_c \wedge A^{[a b}\,j^{c]}_{(A)}-\frac{3}{16}\int d^4x\,\eta_{a b}j_{(A)}^a j_{(A)}^b,
\end{align}
where the last term is the usual four-fermion interaction of the Einstein-Cartan scheme.

Variation with respect to the remaining fields leads to a generalization of the dynamical equations (\ref{dynamical equations}), where now the contribution of fermions is present too, i.e.,
\begin{subequations}\label{dynamical equations 2}
\begin{align}\label{Einstein total equations 2}
& \epsilon^a_{\phantom1b c d}\,e^{b}\wedge\widetilde{R}^{c d} = M^a+\epsilon^a_{\phantom1b c d}\,e^{b}\wedge\left(\widetilde{\omega}^{c}_{\phantom1f}+A^{c}_{\phantom1f}\right)\wedge A^{f d}+\nonumber \\
&+\frac{9}{8}e_{b}\wedge e_{c}\wedge A^{[ab}j^{c]}_{(A)}-\frac{1}{16}\epsilon^{a}_{\ \ bcd}e^{b}\wedge e^{c}\wedge e^{d}\eta_{fg}j^{f}_{(A)}j^{g}_{(A)}+\nonumber \\
&-3\epsilon^{a}_{\ \ bcd}e^{b}\wedge e^{c}\wedge \left[i\overline{\psi}\gamma^d\left(d-\frac{i}{4}\left(\widetilde{\omega}+A\right)\right)\psi-i\left(d+\frac{i}{4}\left(\omega+A\right)\right)\overline{\psi}\gamma^d\psi\right],
\\
&d^{(A)}\star F^{f d} =\epsilon_{a b c}^{\phantom1\phantom1\phantom1[d}\,e^a\wedge e^b\wedge\left(\omega^{c f]}+2A^{c f]}\right)+\frac{27}{4}e^{d}\wedge e^{f}\wedge e_{c}\wedge j^{c}_{(A)}.
\end{align}
\end{subequations}
Consequently, the density of spin of the fermion matter is present in the source term of the Yang-Mills equations for the Lorentz connection fields, and the Einstein equations contain in the rhs not only the energy-momentum tensor of the matter, but also a four-fermion interaction term. The dynamical equations of spinors are formally the same as those of the Einstein-Cartan model with the adjoint of the interaction with the connections of the Lorentz group $A$.

As a result, the Einstein-Cartan contact interaction is recovered in the limit of vanishing Lorentz connections, thus shedding light on the existence of independent connections of the Lorentz group on curved space-time, which modifies profoundly the dynamics of the gravitational field both in absence and in presence of fermion matter. In particular, in the first case, the connections $A$ are in strict relation with the torsion tensor modifying the Riemannian structure of ordinary space-time, while, in the second case, the presence of fermions already modifies the structure of space-time, and the Lorentz connections $A$ contribute to the torsion tensor with a boson term. Moreover, the bosonic and fermionic parts of torsion interact, the latter being a source for the boson part of torsion, and the former the mediator of the interaction between two-fermion torsion terms.

In the most general metric structure, curvature, torsion and non-metricity are present (see for example \cite{rous} for the relation between Riemannian curvature and generalized curvature). In \cite{hehlmac}, the most general parity-conserving quadratic Lagrangian has been established for this metric structure, in terms of the irreducible pieces of non-metricity, torsion and curvature, and a cosmological term is also included.
In the curvature square term, the
irreducible pieces of the symmetric and antisymmetric parts 
 of the curvature 2--form have been introduced. Moreover, Since $V_{\rm
  MAG}$ is required to be an {\em odd} 4--form, if parity conservation
is assumed, it has to be built up with one Hodge star, according to the scheme $F\wedge
\star F$, since the star itself is an odd
operator, as well as the Hilbert-Einstein type term 
 and the cosmological term. Thus $V_{\rm MAG}$ is homogeneous of order one in the star
operator.  This model has also been analyzed in \cite{gag}, where its has been compared with Einstein-Cartan theory, PGT, the implications of the coupling with a scalar field, and the possibility that these models, possibly combined with a suitable symmetry-breaking mechanism, might lead to a consistent quantization program is also envisaged.

The works \cite{rel1,hashi,shira,shirab} (see also \cite{hamm} and the references therein for a review about torsion gravity) deal with models where only torsion and curvature are considered. In particular, the irreducible terms under the action of the Lorentz group are classified as
\begin{subequations}
\begin{align}
&T^\mu=T_{\mu\sigma}^{\ \ \ \ \sigma}\\
&t_{\mu\nu\rho}=T_{\mu\nu\rho}+T_{\nu\rho\mu}-\tfrac{1}{3}\left(T_\nu g_{\mu\rho}+T_\mu g_{\nu\rho}\right)+\tfrac{2}{3}g_{\mu\nu}T_\rho\\
&a_\mu=\tfrac{1}{3}\epsilon_{\mu\nu\rho\sigma}T^{\nu\rho\sigma},
\end{align}
\end{subequations}
i.e. the trace, a traceless part and an antisymmetric part, respectively. The total action accounting for the presence of torsion $L_{tors}$ was taken as
\begin{equation}
L_{tors}=\alpha t_{\mu\nu\rho}t^{\mu\nu\rho}+4\beta T_\mu T^\mu+\gamma a^\mu a_\mu,
\end{equation}
where three constants have been introduced. Furthermore, the procedure was repeated also for the curvature scalar, and parity-violating terms were excluded. When the scalar invariant is added, as a result, a ten-parameter Lagrangian is found.
Our simplified action (\ref{total action reduced}) is forced by the coupling allowed by a geometrical interpretation of this Lorentz gauge theory.
 
\paragraph{Discussion}\null
To better understand the physical implications of first- and second-order approaches, a comparison between field equations should be accomplished. In particular, in the first-order approach, the role of the gravitational filed as a source of torsion should be analyzed, since it has no analogs in GR. Because the gravitational field acts like a source term, it should be compared with a ''current'' term, which can be worked out from the second-order formalism. For the sake of simplicity, we will restrict our analysis to the linearized regime in the transverse-traceless (TT) gauge.

If we consider the case of small perturbations $h_{\mu\nu}$ of a flat Minkowskian metric $\eta_{\mu\nu}$,
\begin{equation}
g_{\mu\nu}=\eta_{\mu\nu}+h_{\mu\nu},
\end{equation}
and the corresponding expression of the tetrad field as a sum of the Minkowskian bein projection $\delta^{a}_{\mu}$ and the infinitesimal perturbation $\zeta^{a}_{\mu}$,
\begin{equation}
\ \ e^{a}_{\mu}=\delta^{a}_{\mu}+\zeta^{a}_{\mu},
\end{equation}
the following identifications hold
\begin{subequations}
\begin{align}
&\eta_{\mu\nu}=\delta_{\mu}^{a}\delta_{a\nu}\\
&h_{\mu\nu}=\delta_{a\mu}\zeta^{a}_{\nu}+\delta_{a\nu}\zeta^{a}_{\mu}.
\end{align}
\end{subequations}
In the linearized regime, all quantities will be truncated at the first order of $\zeta$.

Because of the interaction term (\ref{interacting term}) postulated in the first-order approach, it is possible to solve the structure equation and to express the connections as a sum of the pure gravitational connections plus other contributions, both in absence and in presence of spinor matter. 
The connections $\omega^{ab}_{\ \ \mu}=e^{b\nu}\nabla_{\mu}e^{a}_{\nu}$ rewrite, because of the linearization,
\begin{equation}\label{corre1}
\omega^{ab}_{\ \ \mu}=\delta^{b\nu}\left(\partial_{\nu}\zeta^{a}_{\mu}-\tilde{\Gamma}(\zeta)^{\rho}_{\mu\nu}\delta^{b}_{\rho}\right),
\end{equation}
where $\tilde{\Gamma}(\zeta)^{\rho}_{\mu\nu}$ are the linearized Christoffel symbols, i.e.,
\begin{equation}\label{kris}
\tilde{\Gamma}(\zeta)^{\rho}_{\mu\nu}=\frac{1}{2}\delta^{\rho\sigma}\left(\zeta_{\sigma\mu,\nu}+\zeta_{\sigma\nu,\mu}-\zeta_{\mu\nu,\sigma}\right).
\end{equation}
The Einstein Lagrangian density for $g_{\mu\nu}$ in the TT gauge reads
\begin{equation}
L=\left(\partial_{\rho}h_{\mu\nu}\right)\left(\partial^{\rho}h^{\mu\nu}\right),
\end{equation}
from which $M^{\tau}_{\ \ \alpha\beta}$, the spin-current density associated with the spin angular momentum operator, can be evaluated for a Lorentz transformation of the metric. In fact, if we consider the transformation
\begin{equation}
g_{\mu\nu}\rightarrow\frac{\partial x^{\rho'}}{\partial x^{\mu}}\frac{\partial x^{\sigma'}}{\partial x^{\sigma}}g_{\rho'\sigma'},
\end{equation}
where $x'^{\rho}=x^{\rho}+\epsilon^{\rho}_{\ \ \tau}x^{\tau}$, then the current reads
\begin{equation}\label{corre2}
M^{\tau\phi\upsilon}=\frac{\partial L}{\partial h_{\mu\nu,\tau}}\Sigma^{\rho\phi\upsilon\sigma}_{\mu\nu}h_{\rho\sigma}=\left(\eta^{\chi\mu}\zeta^{\nu,\tau}_{\chi}+\eta^{\chi\nu}\zeta^{\mu,\tau}_{\chi}\right)\Sigma^{\rho\phi\upsilon\sigma}_{\mu\nu}\left(\eta_{\pi\rho}\zeta^{\pi}_{\sigma}+\eta_{\pi\sigma}\zeta^{\pi}_{\rho}\right),
\end{equation}
where
\begin{equation}
\Sigma^{\rho\phi\upsilon\sigma}_{\mu\nu}=\eta^{\chi[\phi}\left(\delta^{\rho}_{\chi}\delta^{\upsilon]}_{\mu}\delta^{\sigma}_{\nu}+\delta^{\rho}_{\mu}\delta^{\sigma}_{\chi}\delta^{\upsilon]}_{\nu}\right).
\end{equation}

The two quantities (\ref{corre1}) and (\ref{corre2}) do not coincide: in fact, (\ref{corre1}) is linear in the $\zeta$ terms, while (\ref{corre2}) is second order in $\zeta$ by construction.
Since the connections $\omega$, on the rhs of (\ref{ymabs}), acquire the physical meaning of a source for torsion, they can be interpreted as a spin-current density. Nevertheless, (\ref{corre1}) is linear in $\zeta$, since the interaction term (\ref{interacting term}) is linear itself; as suggested by the comparison with gauge theories, and with (\ref{corre2}) in particular, the interaction term should be quadratic. In this case, however, it would be very difficult to split up the solution of the structure equation as the sum of the pure gravitational connections plus other contributions.

\section{Concluding Remarks}\label{conclusion}

The considerations developed in this paper have been prompted by observing that GR admits two physically different symmetries, namely the diffeomorphism invariance, defined in the world space-time, and the local Lorentz invariance, associated to the tangent fiber. Such two symmetries reflect the different behaviors of tensors and spinors, respectively, when global Lorentz transformations become local, i.e., while tensors do not experience the difference between the two transformations, spinors do. In the Lorentz-gauge proposal, the diffeomorphism invariance concerns the metric structure of the space-time and it finds in the vier-bein fields the natural gauge counterpart, though the gauge picture holds on a qualitative framework. On the other hand, the real gauge symmetry corresponds to local rotations in the tangent fiber and admits a geometrical gauge field induced by the space-time torsion and its properties. In the present analysis, the keypoint has been fixing the equivalence between isometric diffeomorphisms and local Lorentz transformations. In fact, under, the action of the former, spin connections behave like a tensor and are not able to ensure invariance under the correspondingly-induced local rotations.

This picture can lead one to infer the existence of a (metric-independent) gauge field of the Lorentz group, identified with the connection 1-forms $A$, which interact with spinors. The usual connection 1-forms could not be identified with the suitable gauge fields, for they iare not a primitive object (they depend on bein vectors) and define local Lorentz transformations on the tangent bundle.

The mathematical identification of the Lorentz gauge field with the contortion field follows from the Cartan structure equation if a (unique ) interaction term between the gauge field and the spin connection is introduced. This interaction term induces a Riemannian source to Yang-Mills equations; thus, the real vacuum dynamics of the Lorentz gauge connections takes place on a Minkowski space only, when the Riemannian curvature and the spin currents provide negligible effects, and spin connections can be chosen as vanishing. In fact, it is the geometrical interpretation of torsion as a gauge field that generates the non-vanishing part of Lorentz connections on flat space-time.

Differently from PGT, translations are not treated as a gauge freedom for spinors, because they have no spinor representation, so that spinors are not expected to distinguish between generic diffeomorphisms and translations. Furthermore, the Yang-Mills nature of Lorentz gauge connections enables one to predict propagating fields. Despite these fundamental differences, a pure contact interaction for spinor fields is recovered for vanishing Lorentz connections: in this case, the Cartan structure equation provides non-zero torsion even when gauge bosons are absent. From this point of view, PGT can be qualitatively interpreted as the first-order approximation of the Lorentz-gauge scheme, when the carrier of the interaction is not observable, because of the weakness of its interaction.

As far as short-range interacting torsion is concerned, it is worth remarking that a wide class of models also exists, where the only modes of the torsion tensor which interact with matter are massive scalars. In the valuable analysis presented in \cite{carroll}, for example, the parameter space of the model has been properly constrained, such that torsion decays quickly into matter fields, and no long-range fields are generated, which could be discovered by ground-based or astrophysical experiments. Within this work, in particular, the possibility to introduce propagating torsion and its observational consequences are discussed: possible actions for torsion, consisting in powers and derivatives of torsion, are hypothesized, and the interaction with external matter fields is investigated. Among these possibilities, a single matter-interacting scalar mode is picked up, according to the request that no arbitrary constraints on the dynamics should be present. The main difference between this proposal and our work relies on the choice of a suitable action for torsion and on the motivations for such a choice. In fact, while, in the relevant approach developed in \cite{carroll} several actions for torsion are taken into account, whose implications are evaluated to be not detectable in astro-physical or ground based experiments, our proposal for the action accounting for the presence of torsion is based on attributing a proper  geometrical interpretation to Lorentz connections. 

\section*{Acknowledgments}We would like to thank Simone Mercuri for his advice about these topics.


\begin{thebibliography}{xx}
\bibitem{uti56} 
R. Utiyama, \emph{Phys. Rev.} \textbf{101}, 1597 (1956).

\bibitem{kib61} 
T.W.B. Kibble, \emph{J. Math. Phys.} \textbf{2}, 212 (1961).

\bibitem{blago}
M. Blagojevic, \emph{Gravitation and gauge symmetries}, IoP Publishing (2002). 

\bibitem{hehvdhker76} 
F.W. Hehl, P. von der Heyde, G.D. Kerlick, J. Nester, \emph{Rev. Mod. Phys.} \textbf{48}, 393 (1976).

\bibitem{heh73} 
F.W.Hehl, \emph{Gen. Rel. Grav. J.} \textbf{4}, 333 (1973).

\bibitem{heh74}
F.W. Hehl, \emph{Gen. Rel. Grav. J.} \textbf{5}, 491 (1974).

\bibitem{hehvdhker74}
F.W. Hehl, P. von der Heyde, G.D. Kerlick, \emph{Phys. Rev.} \textbf{D10} 1066 (1974).

\bibitem{hehvdhker74b}
P. Von der Heyde, \emph{Phys. Lett.}, \textbf{A58}, 141 (1976).

\bibitem{re4}
M. Blagojevic, \emph{SFIN} \textbf{A1}, 147 (2003) [gr-qc/0302040].

\bibitem{deis} 
S. Deser, C.J. Isham, \emph{Phys. Rev.}, \textbf{D14}, 2505 (1976).

\bibitem{rel2} 
J. Nitsch, in \emph{Proc. 6th Course of the School of Cosmology and Gravitation on Spin, Torsion and Supergravity}, Erice (Italy) 1979, p.63, eds. P.G.Bergmann and V. de Sabbata, Plenum Press (1980).

\bibitem{rel3} 
F.W. Hehl, in \emph{Proc. 6th Course of the School of Cosmology and Gravitation on Spin, Torsion and Supergravity}, Erice (Italy) 1979, p.5, eds. P.G.Bergmann and V. de Sabbata, Plenum Press (1980).

\bibitem{wal84} 
R. Wald, \emph{General Relativity}, The University of Chicago Press, 1984.

\bibitem{lus1} 
A. Barducci, R. Casalbuoni, L. Lusanna, \emph{Nucl. Phys.} \textbf{B124}, 521 (1976).

\bibitem{neem}
A. Cant, Y. Ne'eman, \emph{J. Math. Phys.} \textbf{26}, 3180 (1985).

\bibitem{neem1}
Y. Ne'eman, D. Sijacki, \emph{Found. Phys} \textbf{27}, 1105 (1997) [gr-qc/9804037].

\bibitem{g2} 
O. M. Lecian, S. Mercuri, in \emph{Proceedings of the 9th Marcel Grossmann meeting} (Berlin 2006), p.2668, World Scientific (2008).

\bibitem{mag2}
F.W. Hehl, J.D. McCrea, E.W. Mielke, Y. Ne'eman, \emph{Phys. Rept.} \textbf{258}, 1 (1995).

\bibitem{obu} 
Y.N. Obukhov, J.G Pereira, \emph{Phys. Rev.} \textbf{D76}, 044016 (2003).

\bibitem{gag}
F. Gronwald, F.W. Hehl, On the Gauge Aspects of Gravity, in \emph{Proc. 14th Course of the School of Cosmology and Gravitation}, Erice (Italy) 1995, World Scientific (1996).

\bibitem{hehlmac} 
F.W. Hehl, A. Macias, \emph{Int. J. Mod. Phys.} \textbf{D8}, 399 (1999).

\bibitem{naka}
N. Carlevaro, O.M. Lecian, G. Montani, \emph{Ann. Fond. L. de Broglie} \textbf{32}, 281 (2007).

\bibitem{nakb}
N. Carlevaro, O.M. Lecian, G. Montani, \emph{Int. J. Mod. Phys.} \textbf{A23}, 1282 (2008).

\bibitem{g1a} 
G. Aprea, G. Montani, R. Ruffini, \emph{Int. J. Mod. Phys. } \textbf{D12}, 1875 (2003).

\bibitem{g1b}
R. de Azeredo Campos, P.S. Letelier, \emph{Progr. Theor. Phys.} \textbf{75}, 1359 (1986).

\bibitem{g1c}
H.J. Xie, T. Shirafuji, \emph{Progr. Theor. Phys.} \textbf{97}, 129 (1997).

\bibitem{hashi} 
K. Hayashi, T. Shirafuji, \emph{Prog. Theor. Phys} \textbf{64}, 866 (1980).

\bibitem{rel1} 
K. Hayashi, T. Shirafuji, \emph{Phys. Rev.} \textbf{D19}, 3524 (1979).

\bibitem{tietre05-2} 
A. Tiemblo, R. Tresguerres, \emph{Eur. Phys. Journ.} \textbf{C42}, 437 (2005).

\bibitem{weisa} 
C. Wiesendanger, in \emph{Proc. on ``Quantum Gravity''}, Erice (Italy) 1995, , World Scientific (1996) [gr-qc/9604043].

\bibitem{weisb} 
C. Wiesendanger, \emph{Class. Quant. Grav} \textbf{13}, 681 (1996).

\bibitem{13} 
K. Hayashi, T. Shirafuji, \emph{Phys. Rev.} \textbf{D19}, 3524 (1979).

\bibitem{14}
F.W. Hehl, J. Nitsch and P. von der Heyde, in \emph{General Relativity and Gravitation - One Hundred Years after the birth of A. Einstein}, Vol.1, p.329, ed. A. Held, Plenum Press (1980).
 
\bibitem{15}
J. Nitsch, in \emph{Cosmology and Gravitation - Spin, Torsion, Rotation and Supergravity}, p.63, eds. P. G. Bergmann and V. de Sabbata, Plenum 1980.
 
\bibitem{16}
M. Blagojevi\'c and M. Vasili\'c, \emph{Quant. Grav.} {\textbf 17}, 3785 (2000).

\bibitem{17a}
W. Kopczy\'nski, \emph{J. Phys.} \textbf{A15}, 493 (1982).

\bibitem{17b}
W-H. Cheng, D-C. Chern and J. M. Nester, \emph{Phys. Rev.} \textbf{D38}, 2656 (1988).
 
\bibitem{17c}
H. Chen, J. M. Nester and H. J. Yo, \emph{Acta Phys. Pol.} \textbf{B29}, 961 (1998).

\bibitem{lecl}
M. Leclerc, \emph{Phys. Rev.} \textbf{D71}, 027503 (2005).

\bibitem{lecl2} 
M. Leclerc, \emph{Phys. Rev.} \textbf{D72}, 044002 (2005).

\bibitem{mandl}
F. Mandl, F. Shaw, \emph{Quantum Field Theory}, J. Wiley and Sons (1984).

\bibitem{weinberg}
S. Weinberg, \emph{The quantum theory of fields}, Cambridge University press (1995).

\bibitem{shap}
I.L. Shapiro, \emph{Phys. Rept.} \textbf{357}, 113 (2002).

\bibitem{oba83} 
T. Obata, \emph{Progr. Theor. Phys.} \textbf{70}, 622 (1983).


\bibitem{lus2} 
F. Bigazzi, L. Lusanna, \emph{Int. J. Mod. Phys.} \textbf{A14}, 1877 (1999).

\bibitem{ash89} 
A. Ashtekar, J.D. Romano, R.S. Tate, \emph{Phys. Rev.} \textbf{D40}, 2572 (1989).

\bibitem{rous} 
S. Casanova, O.M. Lecian, G. Montani, R. Ruffini, R. Zalaletdinov, \emph{Mod. Phys. Lett.} \textbf{A23}, 17 (2008).
 
\bibitem{shira}
K. Hajashi, T. Shirafui, \emph{Prog. Theor. Phys.} \textbf{64}, 883 (1980).

\bibitem{shirab}
K. Hajashi, T. Shirafui, \emph{Prog. Theor. Phys.} \textbf{64}, 1435 (1980).
   
\bibitem{hamm}
R.T. Hammond, \emph{Rep. Prog. Phys.} \textbf{65}, 599 (2002).

\bibitem{carroll} 
S.M. Carroll, G.B. Field, \emph{Phys. Rev.} \textbf{D50}, 3867 (1994).


\end{thebibliography}
\end{document}